\documentclass[twocolumn,10pt]{asme2ej}

\usepackage{color}
\definecolor{light-gray}{gray}{0.80}
\usepackage[]{listings}
\usepackage{xcolor}
\usepackage{amsfonts}
\usepackage{amsmath}
\usepackage{amssymb}
\usepackage{bm}
\usepackage{textcomp,mathcomp}
\usepackage{pgfgantt}
\usepackage{tikz}
\usepackage{caption}
\usepackage{subcaption}
\usepackage[colorlinks,citecolor=red,urlcolor=blue,bookmarks=false,hypertexnames=true]{hyperref} 
\usepackage[linesnumbered,ruled,vlined]{algorithm2e}
\usepackage{algorithmic}
\usepackage{multirow}
\usepackage{graphicx}
\usepackage{lineno}
\usepackage{enumerate}
\usepackage{multirow}
\usepackage{epstopdf}
\usepackage{tikz}
\tikzstyle{startstop} = [rectangle, rounded corners, minimum width=3cm, minimum height=1cm, align=center, text width=4cm, draw=black, fill=red!30 ]
\tikzstyle{io} = [trapezium, trapezium left angle=70, trapezium right angle=110, minimum width=3cm, minimum height=1cm, align=center, text width=4cm, draw=black, fill=blue!30]
\tikzstyle{process} = [rectangle, minimum width=3cm, minimum height=1cm, align=center, text width=4cm, draw=black, fill=orange!30]
\tikzstyle{process1} = [rectangle, minimum width=5cm, minimum height=1cm, align=center, text width=5cm, draw=black, fill=orange!30]
\tikzstyle{decision} = [diamond, minimum width=1cm, minimum height=1cm, align=center, text width=4cm, draw=black, fill=green!30]
\tikzstyle{arrow} = [thick,->,>=stealth]

\graphicspath{{Figures/}}
\title{A phase-field model for thermo-mechanical fracture with an open-source implementation of it using Gridap in Julia}

\author{Ved Prakash
    \affiliation{
    School of Infrastructure\\
	Indian Institute of Technology\\ Bhubaneswar\\
	Odisha-752050, India\\
    Email: vp18@iitbbs.ac.in
    }
}

\author{Akash Kumar Behera
    \affiliation{
    School of Infrastructure\\
	Indian Institute of Technology\\ Bhubaneswar\\
	Odisha-752050, India\\
    Email: s21ce09007@iitbbs.ac.in
    }
}
\author{Mohammad Masiur Rahaman\thanks{Corresponding Author} \\
\affiliation{Assistant Professor,  School of Infrastructure\\
    Indian Institute of Technology\\
	Bhubaneswar\\
    Odisha-752050, India\\
    Email: massiur@iitbbs.ac.in
    }
}

\begin{document}

\maketitle

\begin{abstract}
{\it In this article, we propose a thermodynamically consistent phase-field model for thermo-mechanical fracture and provide an open-source implementation of the proposed model using a recently developed finite element toolbox, Gridap in Julia. Here, we have derived the balance equations for the thermo-mechanical fracture by invoking the virtual power principle and determined the constitutive relations for the thermodynamic fluxes based on the satisfaction of the thermodynamic laws. Our proposed formulation provides an equation of temperature evolution that can easily accommodate dissipative effects such as viscous damping. One may consider the proposed model as a non-trivial extension of a recently developed iso-thermal phase-field model by Dhas {\it{et al.}} \cite{dhas2018phase} for the non-isothermal case. We provide very compact and user-friendly open-source codes for implementing the proposed model using Gridap in Julia that requires very low memory usage and gives a high degree of flexibility to the users in defining weak forms of the governing partial differential equations. We have validated the proposed model and its implementation against such standard results available in the literature as crack propagation in the cruciform shape material, single edge-notched plate, bi-material beam and a quenching test.
}
\end{abstract}


\section{Introduction}
\label{Sec:Introduction}
Analysis of thermally induced stresses and cracks are of great importance in the design of engineering materials. When materials are exposed to a sudden change in ambient temperature, it can result in a non-uniform change of volume that induces thermal stresses and leads to the initiation of cracks. For many practical problems, an iso-thermal model may not be accurate enough to capture the experimental observations when the effect due to the difference between ambient and reference temperature is significant. Specifically, for problems related to thermal shock, materials with inhomogeneous thermal properties, materials subjected to cooling during manufacturing, etc., consideration of thermal effect becomes vital. Some of the practical examples are shrinkage cracks in the concrete, window materials for aircraft \cite{qu2016rapid}, additive manufacturing \cite{gouge2017thermo} etc. Thermally-induced cracks may be complex and may follow paths very different from the one observed for mechanically induced cracks; typically, cracks propagate from higher heating area to lower heating area \cite{chu2017study}. Thermo-mechanical coupling can significantly modify the behavior and can lead to unstable crack growth. Also, understanding the mechanism of thermally induced fracture is essential in the development of novel engineering materials. Hence, fracture analysis of materials under thermo-mechanical loading has become one of the contemporary topics of research interest.

Over the years, several experiments have been carried out to understand the behavior of thermally induced cracks and their effect on the strength or other essential parameters. Geyer and Nemat-Nasser \cite{geyer1982experimental} considered the crack growth behavior of a glass plate where cooling from the boundaries resulted in parallel cracks. They have observed stable crack growth for pre-cooling cracks. The effect of the width of a plate on speed and nature of crack propagation under the presence of thermal gradient was studied by Ronsin {\it{et al.}} \cite{ronsin1995experimental}. Shao {\it{et al.}} \cite{shao2011effect} studied how the residual strength of ceramics varies with the stages of crack propagation. To understand the behavior of infrared window materials used in hypersonic aircraft Qu {\it{et al.}} \cite{qu2016rapid} conducted numerical (finite element analysis) and experimental studies on chemical vapor deposited zinc sulfide (CVD ZnS). Yousef {\it{et al.}} \cite{yousef2005microcrack} experimentally observed the microcrack evolution in large-grained polycrystalline alumina along with the finite element simulation. They also reported variation in macro-structural properties such as young's modulus due to reduction in temperature. To study the stable crack growth in glass Grutzik and Reedy \cite{grutzik2018crack} developed a novel bi-material beam specimen that was tested under temperature cooling conditions. Jiang {\it{et al.}} \cite{jiang2012study} studied the crack patterns in ceramic due to thermal shock both experimentally and numerically using the finite element method.

Modeling of fracture involving realistic conditions is tackled with the help of numerical approaches. Broadly these approaches can be classified as discrete or discontinuous and diffusive or continuous. Discrete methods require modification in the mesh as the fracture process advances, thus making it computationally expensive. The boundary element method (BEM) \cite{prasad1994dual,prasad1996dual} addresses the issue related to mesh by representing the domain by boundaries. Extended finite element method (XFEM) has also been used for modeling the thermoelastic fracture problems \cite{duflot2008extended,zamani2010implementation} that uses additional enrichment functions requiring prior knowledge of the fracture behavior. Meshless approaches \cite{pant2010numerical} have also been used that removes the difficulties involved with the discretization of the domain. Ai and Augarde \cite{ai2019thermoelastic} used the meshless method using adaptive cracking particle method that requires some visibility criteria to model the discontinuities. Recently Greco {\it{et al.}} \cite{greco2021crack} used the moving mesh strategy to model the crack propagation in the FEM framework. Unlike the discrete methods, the diffusive approaches use the effect of fracture on the governing laws, hence eliminating the difficulties associated with modification in the meshing. Gradient damage model, peridynamic and phase-field model, etc., are some of the popular continuum-based approaches. Localizing gradient damage model (LGDM) \cite{poh2017localizing} was used for thermoelastic brittle fracture by Sarkar {\it{et al.}} \cite{sarkar2020thermo}. Recently using the mathematical similarity between damaged material and Riemannian manifold a new continuum-based damage model has been proposed by
Das {\it{et al.}} \cite{das2021geometrically}. Along with these models, some non-local continuum damage \cite{10.1115/1.3173674} studies can also be found in the literature. In this work, we are using the phase-field approach for modeling fracture under thermo-mechanical loading.

Several phase-field models (PFM) have been independently developed in the physics and mechanics community. A brief review can be found in \cite{ambati2015review}. Our focus is on the latter in which the sharp discontinuity (crack) is approximated by a smooth transition of a scalar field variable \emph{s} (known as phase-field) between \emph{0} (completely damaged) to \emph{1} (no damage). In the phase-field models, the brittle fracture is modeled based on the criteria of Griffith \cite{griffith1921vi}. The total energy is considered as a sum of elastic and fracture parts. Based on the similarity of the energy functional with the work of Mumford and Shah \cite{mumford1989optimal} in image segmentation and approximating it as in \cite{ambrosio1990approximation}, Francfort and Marigo \cite{francfort1998revisiting} posed the problem as energy minimization that helped researchers to model fracture problems by a set of governing partial differential equations. Thus, there is no need to redefine the domain, additional enrichment function, etc. The work of Miehe {\it{et al.}} \cite{miehe2010thermodynamically} was extended for multiphysics problems, including the effect of finite deformation thermoelasticity by Miehe {\it{et al.}} \cite{miehe2015phase}. Badnava {\it{et al.}} \cite{badnava2018h} used a mesh refinement algorithm along with the PFM for thermoelastic brittle fracture. To simulate the realistic crack propagation under dynamic loading in a thermoelastic solid Chu {\it{et al.}} \cite{chu2017study} proposed a new elastic energy density and derived the coupling equations based on the energy balance principle. Mandal {\it{et al.}} \cite{mandal2021fracture} extended the work of Wu \cite{wu2017unified} and Wu and Nguyen \cite{wu2018length} for the thermoelastic case including a monolithic BFGS (Broyden–Fletcher–Goldfarb–Shanno) algorithm to solve the coupling equations. Wang {\it{et al.}} \cite{wang2020phase} provided an Abaqus implementation of the thermoelastic problem using explicit time integration, as well as source code that can be used to adapt it to other multifield problems by researchers. The doctoral dissertation of Liu \cite{liu2021phase} shows the capability of the thermo-elastic PFM to simulate fracture in poly-crystalline materials. Hybrid PFM \cite{ambati2015review} was used to model fracture in thick plates by Raghu {\it{et al.}} \cite{raghu2020thermodynamically} using the higher-order shear deformation theory \cite{10.1115/1.3167719}. Along with that, Hybrid PFM has also been used to numerically simulate the thermoelastic brittle fracture by Tangella {\it{et al.}} \cite{tangella2021hybrid}.

In this work, we have developed a new phase-field model for thermo-mechanical brittle fracture. In the proposed model, we have derived the balance equations by invoking the virtual power principle. Using the first and second law of thermodynamics, we have derived the constitutive relations for the thermodynamic fluxes, and the temperature evolution equations, in such a way that any dissipative effects can be easily incorporated into the formulation. One may consider the proposed model as an extension to the phase-field model for brittle fracture, presented by Dhas {\it{et al.}} \cite{dhas2018phase} from iso-thermal to a non-isothermal case. We have provided the details of an open-source implementation of the proposed model for various numerical examples using a newly developed finite element toolbox Gridap  \cite{verdugo2019user,Badia2020,rahaman2021open}, in Julia \cite{bezanson2012julia,Julia-2017}. Julia is a programming language with advantages of simple code formulation as in dynamic languages like \emph{MATLAB} and high performance like static languages, e.g. \emph{C} or \emph{FORTRAN}. Gridap is written in Julia for simulating numerical problems that are governed by partial differential equations. One of the prime advantages of Gridap is, it does not use any back-end support to achieve performance like other FE libraries as in \emph{FEniCS} \cite{alnaes2015archive}. We have organized the rest of this article in the following way. In Section \ref{Sec:PhaseField}, we have derived the coupled governing partial differential equations for modeling fracture under thermo-mechanical loading. In Section \ref{sec:JuliaImplementation}, we have provided the corresponding weak form equations and described the steps involved to implement the proposed PFM using Gridap in Julia. In Section \ref{Sec:NumericalSimulation}, we have shown the capability of the proposed model by reproducing a few standard numerical and experimental results available in the literature. We have finally concluded our article in Section \ref{sec:conclusion}.


\section{Proposed phase-field model}
\label{Sec:PhaseField}
In this section, we provide the details of our proposed model in a small deformation set-up. The proposed model can be considered as a non-trivial extension of a recently developed phase-field model by Dhas {\it{et al.}} \cite{dhas2018phase} from an iso-thermal to a non-isothermal case.

\subsection{Kinematics}
Consider a material body defined by an open set $\Omega$ with boundary $\partial\Omega$ in the three-dimensional Euclidean space $\mathbb{E}^3$. Corresponding to the displacement field, let $\partial\Omega_u$ and $\partial\Omega_t$ be the Dirichlet and Neumann boundaries, respectively. Corresponding to the temperature field, let $\partial\Omega_T$ and $\partial\Omega_q$ be the Dirichlet and Neumann boundaries, respectively. We define the displacement field as
$\boldsymbol{u}:\Omega\rightarrow\mathbb{R}^3$ at any instant of time $t\in\mathrm{R}^{+}$. Within a small deformation set-up, the strain tensor $\boldsymbol{\epsilon}$ is given by
\begin{equation}
\boldsymbol{\epsilon}(\boldsymbol u) = \frac{1}{2}(\nabla \boldsymbol u + \nabla \boldsymbol u^T),
\end{equation}
where $(\cdot)^T$ denotes the transpose of a tensor and $\nabla$ is the gradient operator. For thermo-mechanical problems, $\boldsymbol{\epsilon}$ can be considered to be consisting of an elastic strain  $\boldsymbol{\epsilon}^{\text{elas}}$ and a thermal strain $\boldsymbol{\epsilon}^{\text{th}}$, which can be represented as
\begin{equation}
   \boldsymbol{\epsilon} =  \boldsymbol{\epsilon}^{\text{elas}}  + \boldsymbol{\epsilon}^{\text{th}},
   \label{eq:strainDecomposition}
\end{equation}
where
\begin{equation}
    \boldsymbol{\epsilon}^{\text{th}} = \alpha(T - T_0)\boldsymbol{I},
    \label{eq:ThermalStrain}
\end{equation}
for a temperature change from an initial temperature $T_0$ to current temperature $T$, considering $\alpha$ as the thermal coefficient of the material body with $\boldsymbol I$ denoting the second-order identity tensor.
\begin{figure}[ht!]
 \centering
    \begin{subfigure}[b]{0.49\textwidth}
         \centering
 \includegraphics[width=\textwidth]{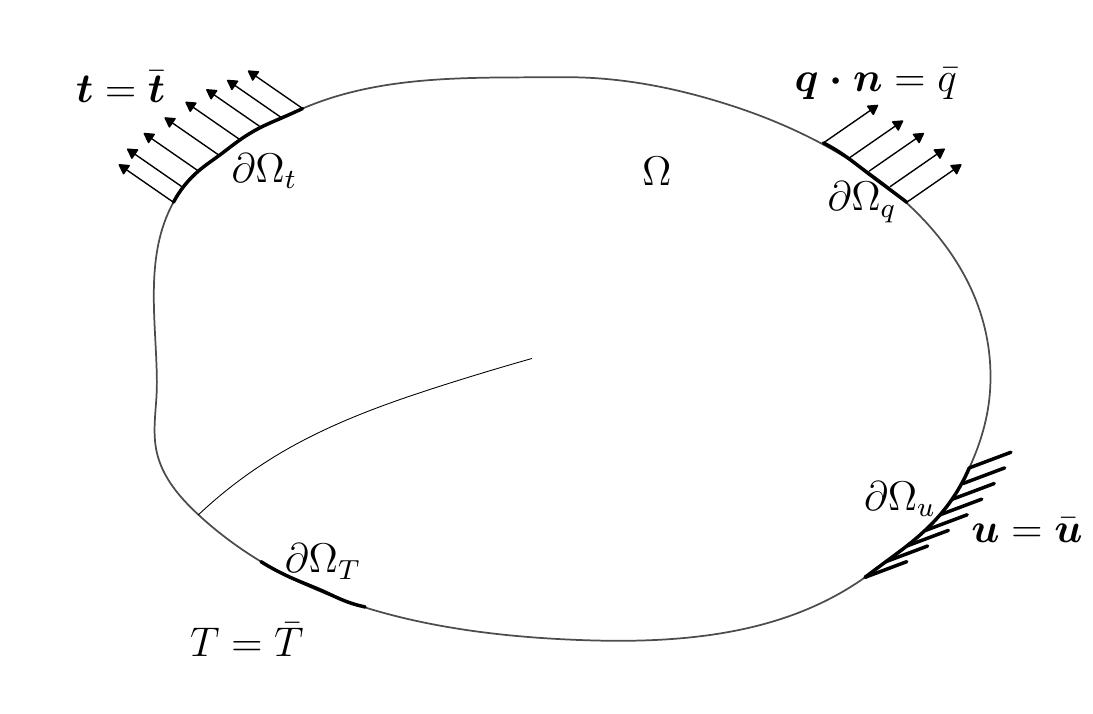}
         \caption{}
     \end{subfigure}
    \begin{subfigure}[b]{0.49\textwidth}
         \centering
 \includegraphics[width=\textwidth]{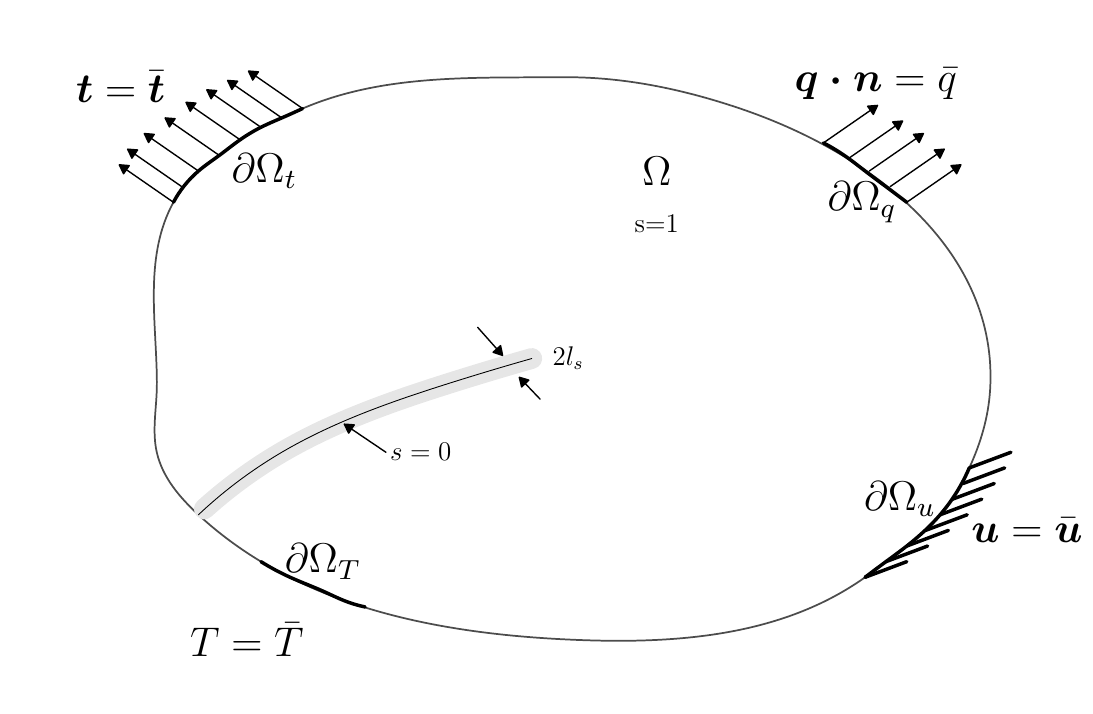}
         \caption{}
     \end{subfigure}
  \caption{A material body is defined by an open set $\Omega$ with Dirichlet boundaries  ($\partial\Omega_T$,$\partial\Omega_u$) and Neumann boundaries ($\partial\Omega_q$,$\partial\Omega_t$) subjected to applied temperature,  displacement, heat flux, and traction $\bar{T}$, $\bar{\boldsymbol{u}}$, $\bar{q}$, $\bar{\boldsymbol{t}}$, respectively. Sub-figure (a) is the material body with sharp crack. For phase-field modeling of fracture a diffused representation of crack is considered in sub-figure (b) by introducing a length scale $l_s$ and a phase-field variable $s$ with $s= 0$ denoting a complete damage, $s = 1$ no damage and $s\in(0,1)$ partial damage.}
\label{fig:DiffDomain}
\end{figure}

In the PFM, sharp cracks are approximated by employing a diffused representation of the cracks via a length scale parameter $l_s$ and a phase-field internal variable $s$ (see Fig. \ref{fig:DiffDomain}). Damaged, undamaged, and partially damaged states of matter can be described using the values of the phase-field variable as follows: $s = 1$ for undamaged state, $s = 0$ for fully damaged state and $s\in(0,1)$ for a partially damaged state. Considering fully damaged states as the fracture, crack set may be defined as $\Omega_s = \{\boldsymbol{x} \in  \Omega|s(\boldsymbol{x}) = 0 \}$. Here, the damage process is considered to be irreversible that is if $\boldsymbol{x} \in \Omega_s$ at time $t_0$ then $\boldsymbol{x} \in \Omega_s$ for all $t \geq t_0$. Thus, the deformed and damaged states of the material body may be described by considering phase-field variable $s$ as an additional kinematic descriptor along with the displacement vector $\boldsymbol{u}$ and the temperature field $T$.

\subsection{Virtual power principle and the balance equations}
  To describe the deformation of a material body under external loading, one can derive the force balances from a virtual power principle by considering a macro-force and a micro-force system  \cite{gurtin1996generalized}. While  stress tensor $\boldsymbol{\sigma}$, traction vector $\boldsymbol{t}(\boldsymbol{n})$ and body force $\boldsymbol{b}$ define the macro-force system, the micro-force system includes scalar micro-traction  $\chi(\boldsymbol{n})$, vector micro-stress $\boldsymbol{\xi}$ and scalar micro-stress $\pi$. It is important to note that the micro-system is also at the same continuum level of the macro-system and the coinage `micro-force' may be a misnomer. One can identify $\dot{\boldsymbol{u}}$ as the power conjugate of  $\boldsymbol{t}(\boldsymbol{n})$ and $\boldsymbol{b}$, $\dot{s}$ as the power conjugate of $\chi(\boldsymbol{n})$ and express the external mechanical power for any arbitrary sub-domain $\mathcal{P}\subset\Omega$ as
  \begin{equation}
     P^{\text{ext}} = \int_{\partial\mathcal{P}}\biggl[\boldsymbol{t} \left(\boldsymbol{n}\right)\cdot \dot {\boldsymbol u} + \chi(\boldsymbol{n}) \cdot \dot s \biggr]\,dA+ \int_\mathcal{P} \boldsymbol{b} \cdot \dot{\boldsymbol{u}} \,dV,
       \label{eq:ExtPower}
  \end{equation}
where, $(\dot{\,\,})$ denotes the time derivative of a variable, $\boldsymbol{n}$ the unit normal vector to the boundary $\partial\mathcal{P}$, $dA$ and $dV$ are, respectively, measures on $\partial\mathcal{P}$ and $\mathcal{P}$. The internal mechanical power $P^{\text{int}}$ can be defined by the summation of power expenditure of $\boldsymbol{\sigma}$ over $\dot {\boldsymbol{\epsilon}}^{\text{elas}}$, $\boldsymbol{\xi}$ over $\nabla \dot {s}$, $\pi$ over $\dot{s}$ and can be given by
    \begin{equation}
     P^{\text{int}} = \int_\mathcal{P} \left(\boldsymbol{\sigma}:\dot {\boldsymbol{\epsilon}}^{\text{elas}} + \boldsymbol{\xi}\cdot\nabla \dot {s} + \pi \,\dot{s} \right) dV.
     \label{eq:IntPower}
    \end{equation}
    We denote the virtual counterparts of $\dot{\boldsymbol{u}}$ and $\dot{s}$ by $\tilde{\boldsymbol{u}}$ and $\tilde{s}$, respectively. Introducing the generalized virtual velocity vector $\mathcal{V} =\left(\tilde{\boldsymbol{u}},\tilde{s}\right)$, one can express the external virtual mechanical power $\tilde{P}^{\text{ext}}\left(\mathcal{P};\mathcal{V}\right)$ and the internal virtual mechanical power $\tilde{P}^{\text{int}}\left(\mathcal{P};\mathcal{V}\right)$ as
    \begin{equation}
    \tilde{P}^{\text{ext}}\left(\mathcal{P};\mathcal{V}\right) = \int_{\partial\mathcal{P}}\biggl[\boldsymbol{t} \left(\boldsymbol{n}\right)\cdot \tilde {\boldsymbol  u} + \chi(\boldsymbol{n}) \cdot  \tilde s \biggr]dA+ \int_\mathcal{P} \boldsymbol{b} \cdot  \tilde{\boldsymbol{ u}} \,dV,
    \label{eq:VirtExtPow}
    \end{equation}
    and
    \begin{equation}
     \tilde{P}^{\text{int}}\left(\mathcal{P};\mathcal{V}\right)= \int_\mathcal{P} \left(\boldsymbol{\sigma}:\tilde {\boldsymbol{\epsilon}}^{\text{elas}} + \boldsymbol{\xi}\cdot\nabla \tilde {s} + \pi \tilde{s} \right) dV,
     \label{eq:VirtIntPow}
    \end{equation}
    respectively. At a given temperature from the symmetry of stress tensor one can show that $\boldsymbol{\sigma}:\tilde{\boldsymbol{\epsilon}}^{\text{elas}} = \boldsymbol{\sigma}:\nabla{\tilde{\boldsymbol{u}}}$ using of which in the virtual power principle, i.e., $ \tilde{P}^{\text{int}}\left(\mathcal{P};\mathcal{V}\right) =  \tilde{P}^{\text{ext}}\left(\mathcal{P};\mathcal{V}\right)$, one can arrive at
    \begin{equation}
    \begin{split}
      \int_{\partial\mathcal{P}}\biggl[\boldsymbol{t} \left(\boldsymbol{n}\right)\cdot \tilde {\boldsymbol  u} + \chi(\boldsymbol{n}) \cdot \tilde s]dA+ \int_\mathcal{P} (\boldsymbol{b} \cdot  \tilde{\boldsymbol{ u}})\,dV =  \int_\mathcal{P} (\boldsymbol{\sigma}:\nabla{\tilde{\boldsymbol{u}}} + \\ \boldsymbol{\xi}\cdot\nabla \tilde {s} + \pi \tilde{s}) \,dV.
       \label{eq:VirtWorkBal}
      \end{split}
    \end{equation}
 Appropriate choices of $\mathcal{V} =\left(\tilde{\boldsymbol{u}},\tilde{s}\right)$ in Eq.  \eqref{eq:VirtWorkBal}  lead to the macro-force and the micro-force balances as described below.
 
 \subsubsection{Macro-force balance}
    \label{subs:Macroscopic_Force_Balance}
Considering $\mathcal{V} =\left(\tilde{\boldsymbol{u}},0\right)$, i.e., by substituting $\tilde{s} = 0 $ in Eq.  \eqref{eq:VirtWorkBal}, and using the divergence theorem the macro-force balance Eq.  may be obtained as
    \begin{equation}
         \int_{\partial\mathcal{P}}\left(\boldsymbol{t} \left(\boldsymbol{n}\right) -\boldsymbol{\sigma}\boldsymbol{n}\right).\tilde {\boldsymbol u}\, dA= -\int_\mathcal{P} \left(\nabla \cdot\boldsymbol{\sigma} + \boldsymbol{b}\right) \cdot \tilde {\boldsymbol{u}}\, dV,
         \label{eq:InvokeTildesZero}
    \end{equation}
which holds for all $\tilde{\boldsymbol{u}}$ and any arbitrary sub-domain $\mathcal{P}$. Applying the localization theorem on Eq.  \eqref{eq:InvokeTildesZero}, one can get that 
    \begin{equation}
        \boldsymbol{t} \left(\boldsymbol{n}\right) = \boldsymbol{\sigma}\,\boldsymbol{n},
    \label{eq:MacroTractionCond}
    \end{equation}
which is the macro-traction condition and
    \begin{equation}
        \nabla \cdot\boldsymbol{\sigma} + \boldsymbol{b} = \boldsymbol{0}.
        \label{eq:MacroforceBal}
    \end{equation}
called the macro-force balance. Equations \eqref{eq:MacroTractionCond} and \eqref{eq:MacroforceBal} may be identified as the classical traction condition and the local linear momentum balance equation, respectively.

\subsubsection{Micro-force balance}
Considering $\mathcal{V} =\left(\boldsymbol{0},\tilde{s}\right)$, i.e., by substituting $\tilde{\boldsymbol {u}} = \boldsymbol{0}$ in Eq.  \eqref{eq:VirtWorkBal}, the equation for micro-force balance can be derived as
      \begin{equation}
         \int_{\partial\mathcal{P}}\biggl[\chi(\boldsymbol{n}) - \boldsymbol{\xi}\cdot \boldsymbol{n}\biggr]\cdot \tilde{s} \,dA = -\int_\mathcal{P} \left(\nabla \cdot \boldsymbol{\xi} - \pi\right)\cdot \tilde{s}  \,dV.
         \label{eq:InvokeVirtTildeuZero}
      \end{equation}
 Since Eq.  \eqref{eq:InvokeVirtTildeuZero} holds for all $\tilde{s}$ and any sub-domain $\mathcal{P}$, Eq.  \eqref{eq:InvokeVirtTildeuZero} may be localized as 
      \begin{equation}
          \chi(\boldsymbol{n}) = \boldsymbol{\xi} \cdot \boldsymbol{n},
     \label{eq:MicroTractionCond}
      \end{equation}
      and 
      \begin{equation}
          \nabla \cdot \boldsymbol{\xi} - \pi = 0.
          \label{eq:MicroforceBal}
      \end{equation}
Equations \eqref{eq:MicroTractionCond} and \eqref{eq:MicroforceBal} are called the micro-traction condition and the micro-force balance, respectively.

\subsection{Thermodynamics laws and constitutive modeling}
 To determine constitutive relations for the thermodynamic fluxes, we have followed the work by Anand and Gurtin
\cite{anand2003thermal}. In presence of heat flux $\boldsymbol{q}$, and heat supply $Q_{\text{h}}$,  the first law of thermodynamics, i.e., the law of conservation of energy, can be given by 
\begin{equation}
\frac{d}{dt} \int_P e\, dV  = P^{\text{int}} - \int_{\partial P}\boldsymbol{q}\cdot \boldsymbol{n}dA + \int_{P}Q_h dV,
\label{eq:RateTotIntEnergy}
\end{equation}
where, $e$ is the internal energy density and the power balance i.e. $P^{ext} = P^{int}$ is used. From the second law of thermodynamics, we have
\begin{equation}
    \frac{d}{dt} \int_P \eta \, dV \ge -\int_{\partial P} \frac{\boldsymbol{q} \cdot \boldsymbol{n}}{T}\, dA + \int_P \frac{Q_{\text{h}}}{T}\, dV,
    \label{eq:RateofEntropy}
\end{equation}
where $\eta$ is the entropy per unit volume. The inequality given by Eq. \eqref{eq:RateofEntropy} ensures that the rate of entropy increase is not less than the entropy input in the system. Using the divergence theorem on Eq.  \eqref{eq:RateTotIntEnergy} and Eq.  \eqref{eq:RateofEntropy}, and applying the localization theorem we get 
\begin{equation}
    \dot{e} =  - \nabla \cdot \boldsymbol{q}  + Q_{\text{h}} + \boldsymbol{\sigma}:\dot{\boldsymbol{\epsilon}}^{\text{elas}} + \boldsymbol{\xi}\cdot\nabla \dot{s} + \pi\,\dot s,
\label{eq:TotIntEnergy}
\end{equation}
and
\begin{equation}
    \dot{\eta} \ge  - \frac{1}{T} \nabla\cdot\boldsymbol{q} + \frac{1}{T^2} \boldsymbol{q} \cdot \nabla T + \frac{Q_{\text{h}}}{T}.
    \label{eq:Entropy}
\end{equation}
From the relation between Helmholtz free energy and internal energy density given by $\psi = e - T\eta$, we have
\begin{equation}
    \dot{\psi} = \dot{e} - \dot{T}\eta - T\dot{\eta}
    \label{eq:RelHelFreeEnIntEn}.
\end{equation}
Employing equations  \eqref{eq:MacroforceBal}, \eqref{eq:MicroforceBal}, \eqref{eq:TotIntEnergy}, \eqref{eq:Entropy} and \eqref{eq:RelHelFreeEnIntEn}, we get that
\begin{equation}
\dot{\psi} -\boldsymbol{\sigma}:\dot{\boldsymbol{\epsilon}}^{\text{elas}} - \boldsymbol{\xi}\cdot\nabla \dot{s} - \pi\,\dot s  + \eta \,\dot T + \frac{1}{T}\boldsymbol{q}\cdot\nabla T \le 0
\label{eq:DisspIneq}.
\end{equation}
The inequality given by Eq. \eqref{eq:DisspIneq} needs to be satisfied for a thermodynamic process to be valid, and one can determine constitutive functions for $\boldsymbol{\sigma}$, $\pi$, $\eta$ and $\boldsymbol{\xi}$ on satisfying Eq. \eqref{eq:DisspIneq}.

\subsection{Constitutive response functions}
\label{Constitutive response functions}
We assume the Helmholtz free energy $\psi$ is a function of $\boldsymbol{\epsilon}^{\text{elas}},\, s,\,\nabla s$, and  $T$, i.e., 
\begin{equation}
    \psi = \hat{\psi}(\boldsymbol{\epsilon}^{\text{elas}},s,\nabla s, T).
\end{equation}
Using the chain rule, the rate of Helmholtz energy can be expressed as
\begin{equation}
    \dot{\psi} = \partial_{\boldsymbol{\epsilon}^{\text{elas}}}\psi : \dot{\boldsymbol{\epsilon}}^{\text{elas}}  + \partial_{s}\psi  \,\dot{s} + \partial_{\nabla s}\psi \cdot {\nabla \dot s} + \partial_{T}\psi\,\dot{T}.
    \label{eq:RatePsi}
\end{equation}
Considering $\pi$ as sum of an energetic part $\pi^{\text{en}}$ and a dissipative part $\pi^{\text{dis}}$, i.e. $\pi = \pi^{\text{en}} + \pi^{\text{dis}}$, and employing Eq. \eqref{eq:RatePsi}, we can rewrite Eq. \eqref{eq:DisspIneq} as
\begin{equation}
\begin{split}
(\boldsymbol{\sigma} - \partial_{\boldsymbol{\epsilon}^{\text{elas}}}\psi) : \dot{\boldsymbol{\epsilon}}^{\text{elas}} +
(\pi^{\text{en}}-\partial_{s}\psi)\dot{s} +  ( \boldsymbol{\xi} - \partial_{\nabla s}\psi ) \cdot {\nabla \dot s} + \\ \pi^{\text{dis}}\dot{s}
- (\eta + \partial_{T}\psi)\dot{T} - \frac{1}{T} \boldsymbol{q}\cdot \nabla T \ge 0.
\label{eq:inequalityMainEquation}
\end{split}
\end{equation}
  Applying Collman and Noll \cite{collman1963thermodynamics} procedure to Eq.  \eqref{eq:inequalityMainEquation}. we obtain the following constitutive relations
   \begin{equation}
          \boldsymbol{\sigma} = \partial_{\boldsymbol{\epsilon}^{\text{elas}}} \psi,
       \label{eq:Constitavestress}
\end{equation}
\begin{equation} \boldsymbol{\xi} = \partial_{\nabla s} \psi,
\label{eq:constitaveVecMicroStress}
\end{equation}    
\begin{equation}
\pi^{\text{en}} = \partial_{ s} \psi,
          \label{eq:constitaveScalarMicroStress}    \end{equation}
      \begin{equation}
          \eta = - \partial_{T} \psi,
     \label{eq:constitutiveEntropy}
      \end{equation}
and the inequality given by \eqref{eq:inequalityMainEquation} reduces to
      \begin{equation}
          \pi^{\text{dis}}\dot{s} - \frac{1}{T} \boldsymbol{q}\cdot \nabla T  \ge 0.
 \label{eq:piDis}
      \end{equation}
      Determination of the constitutive relation must be done in such a way that the inequality constraint given by Eq.  \eqref{eq:piDis} always satisfy. From the irreversiblity condition on damage i.e. $\dot{s}\le 0$, it can be realised that a possible choice for  $\pi^{\text{dis}}$ could be $\pi^{\text{dis}} = \mathcal{G}\dot{s}$, where $\mathcal{G}$ is a constitutive function with $\mathcal{G}\le0$. Now, for the satisfaction of the inequality, the second term also has to be positive. Hence, we define $\boldsymbol{q} = -k\,\nabla T$, where $k$ is the thermal conductivity.
      
\subsection{Specialized constitutive relations}
  We assume the Helmholtz free energy of the system as 
   \begin{equation}
       \psi(\boldsymbol{\epsilon}^{\text{elas}},s,\nabla s, T) = \psi_{\text{elas}}(\boldsymbol{\epsilon}^{\text{elas}}) + \psi_{\text{frac}}(s,\nabla s) + \psi_{\text{ther}}(T),
       \label{eq:Total Energy}
   \end{equation}
   where $\psi_{\text{elas}}$ is the elastic energy, $\psi_{\text{frac}}$ the fracture energy and $\psi_{\text{ther}}$ the thermal energy. For imposing the condition that cracks do not propagate under pure compression, we consider a decomposition of strain tensor into a deviatoric part $\boldsymbol{\epsilon}_{\text{dev}}$ and a volumetric part $\boldsymbol{\epsilon}_{\text{vol}}$, i.e., 
   \begin{equation}
       \boldsymbol{\epsilon} = \boldsymbol{\epsilon}_{\text{dev}} + \boldsymbol{\epsilon}_{\text{vol}},
   \end{equation}
   where
   \begin{equation}
       \boldsymbol{\epsilon}_{\text{dev}} = \mathbb{P}_{\text{dev}}\boldsymbol{\epsilon} ;\,\,\,  \boldsymbol{\epsilon}_{\text{vol}} = \mathbb{P}_{\text{vol}}\boldsymbol{\epsilon}.
       \label{eq:ProjOp}
   \end{equation}
   In Eq. \eqref{eq:ProjOp},  $\mathbb{P}_{\text{dev}}$  and $\mathbb{P}_{\text{vol}}$ projects the strain tensor onto its deviatoric and volumetric parts, respectively. For the assumption of no damage under pure compression, the elastic energy can be expressed as
   \begin{equation}       \psi_{\text{elas}}(\boldsymbol{\epsilon}^{\text{elas}},s) = g(s)\,\psi_{\text{elas}}^+ + \psi_{\text{elas}}^-.
\label{eq:ElastEnerg}
   \end{equation}
In Eq. \eqref{eq:ElastEnerg}, two parts of the elastic energy are given by
   \begin{equation}
       \psi_{\text{elas}}^+ = \frac{1}{2}g(s)\left(\langle p\rangle _+\boldsymbol{I} : \boldsymbol{\epsilon}^{\text{elas}}_{\text{vol}} + \mathbb{P}_{\text{dev}}\mathbb{C} \boldsymbol{\epsilon}^{\text{elas}}: \boldsymbol{\epsilon}^{\text{elas}}_{\text{dev}}\right),
    \label{eq:PosElastoMechenergy}
    \end{equation}
    and
    \begin{equation}
       \psi_{\text{elas}}^- = \frac{1}{2}\langle p\rangle_-\boldsymbol I: \boldsymbol{\epsilon}^{\text{elas}}_{\text{dev}},
    \label{eq:NegElastoMechenergy} 
    \end{equation}
 where $\mathbb{C}$ is the undamaged fourth order elasticity tensor, $\boldsymbol{\epsilon}^{\text{elas}}_{\text{vol}} = \mathbb{P}_{\text{vol}}\, \boldsymbol{\epsilon}^{\text{elas}}$, $\boldsymbol{\epsilon}^{\text{elas}}_{\text{dev}} = \mathbb{P}_{\text{dev}} \,\boldsymbol{\epsilon}^{\text{elas}}$ and g(s) is a degradation function considered as $s^2$ in this work, $\langle p\rangle _+ = \frac{1}{2}\left(p+\lvert p \rvert \right)  $ \,\, and\,\,  $\langle p\rangle _- = \frac{1}{2}\left(p-\lvert p
 \rvert \right) $,\,\, $p = \frac{1}{3}\boldsymbol{I}:\mathbb{P}_{\text{vol}}\mathbb{C}\boldsymbol{\epsilon}^{\text{elas}} $. The fracture energy $\psi_{\text{frac}}(\nabla s,s)$ may be postulated as
   \begin{equation}
       \psi_{\text{frac}}\left(\nabla s,s\right) = G_c\biggl(\frac{\left(1-s\right)^2}{2l_s} + \frac{l_s}{2}\nabla s \cdot  \nabla s\biggr),
      \label{eq:fracEnergy}
   \end{equation}
    where $G_c$ is the critical energy release rate and $l_s$ is the phase-field length scale parameter.
Recognizing the fact that the heat flux may get affected by crack generation we express the thermal conductivity $k$ as
\begin{equation}
    k = g(s)\,k_0,
    \label{eq:SpecialisedHeatConductivity}
\end{equation}
where $k_0$ is the undamaged thermal conductivity and the expression given by Eq. \eqref{eq:SpecialisedHeatConductivity} can account for reduced thermal conduction due to fracture considering the adiabatic fracture zone. With the specialized total energy, one can find out the expressions for thermodynamic fluxes by using the expressions from equations  \eqref{eq:Constitavestress}, \eqref{eq:constitaveVecMicroStress} and \eqref{eq:constitaveScalarMicroStress} as
 \begin{equation}
     \boldsymbol{\sigma} =  g(s)\left(\langle p\rangle_+\boldsymbol{I} + \mathbb{P}_{\text{dev}}\mathbb{C}\boldsymbol{\epsilon}^{\text{elas}}\right) + \langle p\rangle_- \boldsymbol{I},
     \label{eq:sigmaConst}
 \end{equation}
 \begin{equation}
          \boldsymbol{\xi} = G_cl_s\nabla s,
          \label{eq:vectarmicrostress}
 \end{equation}
 \begin{equation}
 \begin{split}
     \pi^{en} = g'(s)\left(\langle p\rangle _+\boldsymbol{I}:\boldsymbol{\epsilon}^{\text{elas}}_{\text{vol}} + \mathbb{P}_{\text{dev}}\mathbb{C} \boldsymbol{\epsilon}^{\text{elas}}: \boldsymbol{\epsilon}^{\text{elas}}_{\text{dev}}\right) -\\ \frac{G_c}{l_s}(1-s).
     \label{eq:scalarmicrostress}
\end{split}
 \end{equation}
 Using the expression for the above thermodynamic fluxes in Eq.  \eqref{eq:MacroforceBal} and Eq.  \eqref{eq:MicroforceBal}, one can arrive at the governing equation for displacement field and phase-field, respectively.
 
\subsection{Temperature evolution}
To derive the temperature evolution equation, we are following the approach similar to the one adopted by Rahaman {\it{et al.}} \cite{rahaman2017dynamic} in a small deformation set-up with the presence of fracture energy instead of energy contribution due to plasticity. Employing Eq. \eqref{eq:constitutiveEntropy} and Eq. \eqref{eq:Total Energy} in the relation $ e = \psi + T\eta$,  we get that 
\begin{equation}
    e = \psi_{\text{elas}} + \psi_{\text{frac}} + \psi_{\text{ther}} - T\frac{d\psi_{\text{ther}}}{dT}.
    \label{eq:IntEn}
\end{equation}
For the case of no mechanical strain, differentiation of Eq. \eqref{eq:IntEn} leads to
\begin{equation}
   de = -T\frac{d^2\psi_{\text{ther}}}{dT^2}dT = \rho\,c\,dT,
    \label{eq:DifferentialEntropy}
\end{equation}
where $c$ is the specific heat. Employing Eq. \eqref{eq:IntEn} and Eq. \eqref{eq:DifferentialEntropy}, we get that
\begin{equation}
    \dot{e} = \dot\psi_{\text{elas}} + \dot\psi_{\text{frac}} + \rho c\dot{T}.
    \label{eq:Enthalpy}
\end{equation}
Expressing $\dot{\psi}_{\text{elas}}$ and $\dot{\psi}_{\text{frac}}$ in their unspecialized form as
\begin{equation}
    \dot{\psi}_{\text{elas}} = \boldsymbol{\sigma}:\dot{\boldsymbol{\epsilon}}^{\text{elas}},
    \label{eq:UnspecilizedElasticEnergy}
\end{equation}
\begin{equation}
    \dot{\psi}_{\text{frac}} = \boldsymbol{\xi}\cdot\nabla \dot s + \pi^{\text{en}} \dot s,
    \label{eq:UnspecilizedFractureEnergy}
\end{equation}
and using Eq.  \eqref{eq:TotIntEnergy} in Eq.  \eqref{eq:Enthalpy} we get the temperature evolution equation as given by
\begin{equation}
   \rho c\dot{T} =  -\nabla\cdot \boldsymbol{q} + Q_{\text{h}} + \pi^{\text{dis}}\dot s.
    \label{eq:TempStrongForm}
\end{equation}
As can be seen from Eq. \eqref{eq:TempStrongForm}, our proposed formulation provides a thermodynamically consistent accommodation of dissipative effects such as temperature change due to viscous damping through the $\pi^{\text{dis}}\dot s$ in Eq.  \eqref{eq:TempStrongForm}. 

  \subsection{Boundary value problem and the strong form}
      \label{sec:BVPdef}
     Using the expressions derived in the previous section, one may write the strong form of the governing PDEs as a boundary value problem and derive the corresponding weak form of the governing PDEs for the finite element formulation of the proposed phase-field model. One can express the stress tensor $\boldsymbol{\sigma}$ in terms of kinematic variables as $\mathbb{C}_{\text{mod}}\, \boldsymbol{\epsilon}^{\text{elas}}$ by defining $\mathbb{C}_{\text{mod}}$ as
\begin{equation}
 \mathbb{C}_\text{mod} = 
 \begin{cases}
   g(s) \mathbb{C},\,\,\,\,\,\text{for}\,\,p \ge 0  & \\
    g(s)\left(\mathbb{P}_{\text{dev}}\mathbb{C}\right) + \mathbb{P}_{\text{vol}}\mathbb{C}, \,\,\,\,\text{for}\,\,p < 0.
  \end{cases}
  \label{eq:CtensorMod}
 \end{equation}
Using $\boldsymbol{\sigma} = \mathbb{C}_{\text{mod}}\, \boldsymbol{\epsilon}^{\text{elas}}$, Eq.  \eqref{eq:MacroforceBal} can be re-written as
\begin{equation}
\nabla \cdot\biggl(\mathbb{C}_{\text{mod}}\, \boldsymbol{\epsilon}^{\text{elas}}\biggr) + \boldsymbol{b} = \boldsymbol{0}.
     \label{eq:ElastStrongForm}
      \end{equation}
Similarly, using Eq.  \eqref{eq:MicroforceBal}, \eqref{eq:vectarmicrostress} and \eqref{eq:scalarmicrostress}, we can re-write the force balance corresponding to phase field as
  \begin{equation}
  \begin{split}
      \nabla\cdot( G_cl_s\nabla s ) - g'(s)\left(\langle p\rangle _+\boldsymbol{I} : \boldsymbol{\epsilon}^{\text{elas}}_{\text{vol}} + \mathbb{P}_{\text{dev}}\mathbb{C} \boldsymbol{\epsilon}^{\text{elas}}: \boldsymbol{\epsilon}^{\text{elas}}_{\text{dev}}\right)\\ + \frac{G_c}{l_s}\left(1-s\right) + \mathcal{G}\dot{s} = 0.
          \end{split}
          \label{eq:RawFracStrongForm}
      \end{equation}
As it can be seen from the above equation, any dissipative effect can easily be incorporated in to the equation through the term $\mathcal{G}\dot s$, bypassing the empirical route by Miehe {\it{et al.}} \cite{miehe2010thermodynamically}. However, in the numerical examples of the present work, we have not considered any energy dissipation and implementation of this model for problems with dissipative effect can be considered as a future scope of the current work. To account for the irreversiblity condition on damage, a history function $\mathcal{H}(\mathcal{E})$, where  $\mathcal{H}(f) = \text{max}_{\tau \in [0,t]}f(\tau)$ for any input argument $f$, of the so-called positive part of elastic energy $\mathcal{E} =\psi^+_{\text{elas}}(\boldsymbol{\epsilon}^{\text{elas}})$ is employed \cite{miehe2010phase}. Using the history function $\mathcal{H}(\mathcal{E})$, Eq.  \eqref{eq:RawFracStrongForm} may be expressed as
\begin{equation}
          \nabla \cdot \biggl(G_cl_s\nabla s\biggr) - g'(s)\mathcal{H}(\mathcal{E})+\frac{G_c}{l_s}\left(1-s\right) = 0,
          \label{eq:FracStrongForm}
      \end{equation}
      where 
      \begin{equation}
          \mathcal{H}(\mathcal{E}) = \langle p\rangle _+\boldsymbol{I}:\boldsymbol{\epsilon}^{\text{elas}}_{\text{vol}} + \mathbb{P}_{\text{dev}}\mathbb{C} \boldsymbol{\epsilon}^{\text{elas}}: \boldsymbol{\epsilon}^{\text{elas}}_{\text{dev}}. 
     \label{eq:HistoryFun}
  \end{equation}
For $\pi^{\text{dis}} =0$ and $\boldsymbol{q} = -k\,\nabla T$, the temperature evolution equation given by Eq. \eqref{eq:TempStrongForm} can be re-written as 
\begin{equation}
    \rho c\dot{T} - \nabla \cdot (k\,\nabla T) = Q_{\text{h}}.
    \label{eq:TempStrongFormFinal}
\end{equation}
The governing PDEs \eqref{eq:ElastStrongForm}, \eqref{eq:FracStrongForm} and \eqref{eq:TempStrongFormFinal} are subjected to boundary conditions, such as  prescribed displacement $\boldsymbol{u} = \bar{\boldsymbol{u}}$, prescribed temperature $T = \bar{T}$ and applied traction $\boldsymbol{t}(\boldsymbol{n}) = \bar{\boldsymbol{t}}(\boldsymbol{n})$ on $\partial\Omega_u$, $\partial\Omega_{T}$, $\partial\Omega_t$, respectively. Equations \eqref{eq:ElastStrongForm}, \eqref{eq:FracStrongForm}, and \eqref{eq:TempStrongFormFinal} together with the boundary conditions are called the strong form of the governing PDEs. For instance, in absence of body force i.e. $\boldsymbol{b}=\bf{0}$, one possible strong form can be given as
 \begin{subequations}
 \begin{align}
     \nabla \cdot \boldsymbol{\sigma} = {\bf{0}} \,\,\,\,\,\left(\text{where}\,\,\boldsymbol{\sigma} = \mathbb{C}_{\text{mod}}\, \boldsymbol{\epsilon}^{\text{elas}}\right)\,\,\,\,\text{in}\,\,\,\,\Omega,\\
\boldsymbol{u} = \bar{\boldsymbol{u}} \,\,\,\,\text{on}\,\,\,\,\partial\Omega_u,\\
 \boldsymbol{\sigma}\,\boldsymbol{n} = \boldsymbol{0}\,\,\,\,\text{on}\,\,\,\,\partial\Omega\backslash\partial\Omega_u,
\end{align}
\label{eq:StrongFormDisplacement}
\end{subequations}
 corresponding to the macro-force balance,
 \begin{subequations}
 \begin{align}
    \nabla \cdot \biggl(G_cl_s\nabla s\biggr) - 2s\,\mathcal{H}(\mathcal{E})+\frac{G_c}{l_s}\left(1-s\right) = 0\,\,\,\,\text{in}\,\,\,\,\Omega,\\
    \nabla s \cdot\boldsymbol{n} = 0 \,\,\,\,\text{on}\,\,\,\,\partial\Omega\backslash\partial\Omega_u,
 \end{align}
 \label{eq:StrongFormPhaseField}
\end{subequations}
corresponding to the micro-force balance, and
\begin{subequations}
 \begin{align}
     \rho \,c\,\dot{T} - \nabla \cdot (k\,\nabla T) = Q_{\text{h}}\,\,\,\,\text{in}\,\,\,\,\Omega,\\
 {T} = \bar{{T}} \,\,\,\,\text{on}\,\,\,\,\partial\Omega_{T},\\     
  \boldsymbol{q}\cdot\boldsymbol{n} =0\,\,\,\,\text{on}\,\,\,\,\partial\Omega\backslash\partial\Omega_T,
 \end{align} 
 \label{eq:TempStrongFormWithBC}
\end{subequations}
corresponding to the temperature evolution. Considering isotropic materials for the present study, we express the components of the fourth order elasticity tensor as $\mathbb{C}_{ijkl} = \lambda\delta_{ij}\delta_{kl} + \mu(\delta_{ik}\delta_{jl}+\delta_{il}\delta_{jk})$, where $\lambda$ and $\mu$ are the Lam\'e parameters and $\delta$ denotes the Kronecker delta. The Lam\'e parameters are related to Young's modulus $E$ and Poisson's ratio $\nu$ by $\lambda = E\nu/\left((1 + \nu)(1 - 2\nu)\right)$ and $\mu = E/\left(2(1 + \nu)\right)$.

\section{The weak form and an outline for finite element implementation in Julia}
  \label{sec:JuliaImplementation}
  
  In this section, we have derived the weak form for the strong form given by equations \eqref{eq:StrongFormDisplacement}, \eqref{eq:StrongFormPhaseField} and \eqref{eq:TempStrongFormWithBC} and outlined the finite element formulation through a numerical example. Let $\boldsymbol{v}$, $\phi$ and $w$ be the test functions corresponding to displacement $\boldsymbol{u}$, phase-field $s$, and temperature field $T$, respectively. The trial spaces with the given displacement and temperature boundary condition may be given by
\begin{equation}
          H_u = \{\boldsymbol{u} \in H^1(\Omega)^d ; \; \boldsymbol{u} = \bar{\boldsymbol{u}} \; \text{on} \; \partial\Omega_u \},
     \label{eq:TrialSpaceU}
      \end{equation}
    \begin{equation}
        H_s = \{s \in H^1(\Omega) \},
    \end{equation}
\begin{equation}
          H_{T} = \{T \in H^1(\Omega) ; \; T= \bar{T} \; \text{on} \; \partial\Omega_{T}\},
    \label{eq:TrialSpaceT} \end{equation}
where $\bar{\boldsymbol{u}}$ and $\bar{T} $ are prescribed displacement and temperature on $\partial\Omega_u$ and $\partial\Omega_{T}$, respectively, and $d$ corresponds to the dimension of displacement vector. The test spaces may be defined as
    \begin{equation}
          V_u = \{\boldsymbol{v} \in H^1(\Omega)^d; \; \boldsymbol{v} = 0 \; \text{on} \; \partial\Omega_u \},
    \label{eq:TestSpaceU}
    \end{equation}
    \begin{equation}
          V_s = \{\phi \in H^1(\Omega) \},
    \end{equation}
    \begin{equation}
    V_{T} = \{w \in H^1(\Omega); \; w = 0 \; \text{on} \; \partial\Omega_{T}\}.
    \label{eq:TestSpaceT}
    \end{equation}
  One can obtain the following weak form: Find $\boldsymbol{u}\in H_u$,  $s\in H_s$ and $T\in H_{T}$ such that for all $\boldsymbol{v}\in V_u$, $\phi\in V_s$ and $w\in V_{T}$
 \begin{subequations}
 \begin{align}
     a_{\text{Disp}}((\boldsymbol{u},T),(\boldsymbol{v},w)) = b_{\text{Disp}}(\boldsymbol{v},w),\\
    a_{\text{PF}}(s,\phi) = b_{\text{PF}}(\phi)
 \end{align}
 \label{eq:WeakForm}
\end{subequations}
where
 \begin{subequations}
 \begin{gather}
 \begin{split}
a_{\text{Disp}}((\boldsymbol{u},T),(\boldsymbol{v},w)) = \int_\Omega( \boldsymbol{\epsilon}(\boldsymbol{v}):\boldsymbol{\sigma}_T(\boldsymbol{u},T) + w \,\rho\, c \,\dot{T} \\ + \nabla w \cdot (k \,\nabla T) 
)\, d{\Omega},
\end{split}
\\
 b(\boldsymbol{v},w)=\int_\Omega\left( \boldsymbol{\epsilon}(\boldsymbol{v}):\boldsymbol{\sigma}_F + Q_{\text{h}}\right)\, d{\Omega},\\ 
a_{\text{PF}}(s,\phi)= \int_\Omega\bigl(G_c\,l_s\,\nabla s \cdot \nabla \phi + 2s\,\phi\,\mathcal{H} +   \frac{G_c}{l_s}\,s \,\phi\bigr)\,d\Omega, \\    b_{\text{PF}}(\phi)= \int_\Omega\frac{G_c}{l_s}\phi d{\Omega},
  \end{gather}
 \label{eq:WeakFormParts}
\end{subequations}
with the definitions of
 $\boldsymbol{\sigma}_T(\boldsymbol{u},T)$ and  $\boldsymbol{\sigma}_F$ as
\begin{subequations}
 \begin{gather}
 \boldsymbol{\sigma}_T(\boldsymbol{u},T) = g(s)\left(\langle p\rangle_+\boldsymbol{I} + \mathbb{P}_{\text{dev}}\mathbb{C}(\boldsymbol{\epsilon}-\alpha\boldsymbol{I} T)\right) + \langle p\rangle_- \boldsymbol{I},\\
  \boldsymbol{\sigma}_F = -g(s)\left(\langle p\rangle_+\boldsymbol{I} + \mathbb{P}_{\text{dev}}\mathbb{C}\alpha\boldsymbol{I} T_0\right) + \langle p\rangle_- \boldsymbol{I}.
 \end{gather}
 \label{eq:StressTotAndThermal}
\end{subequations}
For numerical implementation of the proposed model, we have adopted the staggered approach  proposed by Miehe {\it{et al.}} \cite{miehe2010phase}. At each step, we have solved for the temperature and the displacement field from the couple set of equations \eqref{eq:TempStrongFormFinal} and  \eqref{eq:StrongFormDisplacement} using the MultiFieldFESpace available in gridap. We have computed the solution for the phase field variable using Eq. \eqref{eq:StrongFormPhaseField} based on the history function given by Eq. \eqref{eq:HistoryFun}.
      
One of the most appealing aspects of Gridap is that it allows one to use the weak form directly for finite element implementation. We describe all the steps associated with finite element simulations for the proposed model, such as creating the mesh file, implementing the weak form, applying boundary conditions, solving for unknown field variables, and the post-processing through a Julia code for numerical simulation of brittle fracture in a single edge notched plate under thermo-mechanical loading (see Section \ref{sec:SENT}). One can easily modify the Julia code for simulating other thermo-mechanical fracture problems. To replicate the results reported in Section (\ref{Sec:NumericalSimulation}), as a first step one needs to import the Julia packages listed below into a script file currently created in a jupyter notebook.
\begin{lstlisting}[language=C,backgroundcolor=\color{gray!5!white},showstringspaces=false,breaklines=true,basicstyle=\ttfamily]
using GridapGmsh
using Gridap
using Gridap.Geometry
using Gridap.TensorValues
using PyPlot
\end{lstlisting}
One can define parameters related to applied boundary conditions for a single edge notched plate under thermo-mechanical loading by the following lines in Julia.
\begin{lstlisting}[language=C,backgroundcolor=\color{gray!5!white},showstringspaces=false,breaklines=true,basicstyle=\ttfamily]
const T0 = 300
const TAppMax = T0 + 50
const delt = 1e-2
const tMax = 1
const uMax = 1.2e-6
AppVel = uMax/tMax
uMin = 0
uTran = 0.2*uMax
\end{lstlisting}
We have imposed the prescribed temperature on Dirchlet boundary $\partial\Omega_T$ in small increments till the temperature reaches at the maximum applied temperature and kept at constant maximum applied temperature till the failure of the specimen. For non-linear smooth increment of temperature, one may use the following function.   
\begin{lstlisting}[language=C,backgroundcolor=\color{gray!5!white},showstringspaces=false,breaklines=true,basicstyle=\ttfamily]
using SymPy
x,$x_1$ = symbols("x,$x_1$", real = true)
heaviside(x) = 0.5 * (sign(x) + 1)
interval(x, a, b) = heaviside(x-a) - heaviside(x-b)
hS = uMax/10
F(x) = (T0 - TAppMax) * interval(x,-4*hS+uMin,uTran)
$w_h$(x,$x_1$) = (1/(sqrt(2*pi)*hS))*exp(-(x-$x_1$)^2/(2*hS^2))
smoothT = SymPy.integrate(F($x_1$)*$w_h$(x,$x_1$),($x_1$,-4*hS,uMax)) +  TAppMax
plot(smoothT,0,uMax)
\end{lstlisting}
One may also increase temperature linearly from initial temperature to the maximum applied temperature based on the convergence requirements using the below function.
\begin{lstlisting}[language=C,backgroundcolor=\color{gray!5!white},showstringspaces=false,breaklines=true,basicstyle=\ttfamily]
function Tfun(u)  
    if u <= uTran
      return ((TAppMax - T0)/uTran)*u + T0
    else
     return  TAppMax
    end
end 
plot(Tfun,0,uMax)
\end{lstlisting}
In this work, we have used a non-linear smooth increment of temperature by choosing option one of the two functions as given below.
\begin{lstlisting}[language=C,backgroundcolor=\color{gray!5!white},showstringspaces=false,breaklines=true,basicstyle=\ttfamily]
uAppVec = range(0,uMax,length = Int64(floor(tMax/delt)))

AppTOption = 1 ## 1 for smooth and otherwise linear than constant
if AppTOption == 1
    TAppVec = smoothT.(uAppVec)
  else
    TAppVec = Tfun.(uAppVec) 
end 
\end{lstlisting}
Input material parameters related to elastic, fracture and thermal properties of the notched plate can be defined by the following lines in Julia.
\begin{lstlisting}[language=C,backgroundcolor=\color{gray!5!white},showstringspaces=false,breaklines=true,basicstyle=\ttfamily]
const E = 340.0e9
const $\nu$ = 0.22

const Gc = 42.47
const lsp = 3.33e-6
const $\eta$ = 1e-8

const $\alpha$ = 8e-6
const c = 0.775
const $\kappa$_mat = 300.0
const $\rho$ = 2450

\end{lstlisting}
For the finite element simulations, one needs to have the discretization of the computational domain that can be generated by writing a mesh file ``SquarePlateWithEdgeNotch.msh" in Julia (see \href{https://github.com/MdMasiurRahaman/PhaseFieldModelForThermoMechanicalFracture-/blob/main/MeshNotchedPlateUnderTension.ipynb}{Julia code for meshing of single edge notched plate}) and loading that mesh file to build an instance of ``DiscreteModel" by the following lines.
\begin{lstlisting}[language=C,backgroundcolor=\color{gray!5!white},showstringspaces=false,breaklines=true,basicstyle=\ttfamily]
model = GmshDiscreteModel("SquarePlateWithEdgeNotch.msh")
writevtk(model,"SquarePlateWithEdgeNotch")
\end{lstlisting}
We have defined a function for the constitutive matrix of isotropic material under plane stress or plane strain condition by the following lines in Julia.
\begin{lstlisting}[language=C,backgroundcolor=\color{gray!5!white},showstringspaces=false,breaklines=true,basicstyle=\ttfamily]
function ElasFourthOrderConstTensor($E$,$\nu$,PlanarState)
    # 1 for Plane Stress and 2 Plane Strain Condition 
  if PlanarState == 1
      C1111 = $E$/(1-$\nu$*$\nu$)
      C1122 = ($\nu$*$E$)/(1-$\nu$*$\nu$)
      C1112 = 0.0
      C2222 = $E$/(1-$\nu$*$\nu$)
      C2212 = 0.0
      C1212 = $E$/(2*(1+$\nu$))
  elseif PlanarState == 2
      C1111 = ($E$*(1-$\nu$*$\nu$))/((1+$\nu$)*(1-$\nu$-2*$\nu$*$\nu$))
      C1122 = ($\nu$*$E$)/(1-$\nu$-2*$\nu$*$\nu$)
      C1112 = 0.0
      C2222 = ($E$*(1-$\nu$))/(1-$\nu$-2*$\nu$*$\nu$)
      C2212 = 0.0
      C1212 = $E$/(2*(1+$\nu$))
  end
  C_ten = SymFourthOrderTensorValue(C1111,C1112,C1122,C1112,C1212,C2212,C1122,C2212,C2222)
    return  C_ten
end
\end{lstlisting}
For a plane strain condition, the constitutive tensor can be computed by calling the above Julia function as given below.
\begin{lstlisting}
const C_mat = ElasFourthOrderConstTensor(E,$\nu$,2)
\end{lstlisting}
To satisfy the assumption that crack can not propagate under pure compression, stress and strain tensors are decomposed into a volumetric and a deviatoric part by introducing the projection operators $\mathbb{P}_{\text{vol}}$ and $\mathbb{P}_{\text{dev}}$, respectively, which are defined by the following lines.
\begin{lstlisting}
I2 = SymTensorValue{2,Float64}(1.0,0.0,1.0)
I4 = I2 $\otimes$ I2
I4_sym = one(SymFourthOrderTensorValue{2,Float64})
P_vol = (1.0/3)*I4
P_dev = I4_sym - P_vol
\end{lstlisting}
To express the stress tensor in terms of kinematics variables, i.e.,  $\boldsymbol{\sigma}=\mathbb{C}_{\text{mod}}\,\boldsymbol{\epsilon}$, where the expression of $\mathbb{C}_{\text{mod}}$ is given by Eq.  \eqref{eq:CtensorMod}, the following function is defined in Julia. 
\begin{lstlisting}
$\sigma$_elas($\varepsilon$Elas) = C_mat $\odot$ $\varepsilon$Elas

 function $\sigma$_elasMod($\varepsilon$, $\varepsilon$_in, s_in,T,T_in)
 $\varepsilon$Elas_in = $\varepsilon$_in - $\alpha$*(T_in-T0)*I2
   $\varepsilon$Elas = $\varepsilon$ - $\alpha$*(T-T0)*I2
if tr($\varepsilon$Elas_in)  >= 0
      $\sigma$ = (s_in^2 + $\eta$)*$\sigma$_elas($\varepsilon$Elas)
  elseif tr($\varepsilon$Elas_in) < 0
      $\sigma$ = (s_in^2 + $\eta$) *P_dev $\odot$ $\sigma$_elas($\varepsilon$Elas) + P_vol$\odot$ $\sigma$_elas($\varepsilon$Elas) 
  end  
    return $\sigma$
end
\end{lstlisting}
The function for $\boldsymbol{\sigma}_T$ given by Eq.  \eqref{eq:StressTotAndThermal} can be defined as given below.
\begin{lstlisting}
function $\sigma$_totMod($\varepsilon$, $\varepsilon$_in,s_in,T,T_in)
$\varepsilon$Elas_in = $\varepsilon$_in -$\alpha$*(T_in-T0)*I2
   $\varepsilon$ElasTot = $\varepsilon$ -$\alpha$*T*I2
if tr($\varepsilon$Elas_in)  >= 0
      $\sigma$T = (s_in^2 + $\eta$)*$\sigma$_elas($\varepsilon$ElasTot)
  elseif tr($\varepsilon$Elas_in) < 0
      $\sigma$T = (s_in^2 + $\eta$) *P_dev $\odot$ $\sigma$_elas($\varepsilon$ElasTot) + P_vol$\odot$ $\sigma$_elas($\varepsilon$ElasTot) 
  end  
    return $\sigma$T
end
\end{lstlisting}
Similarly, the function for $\boldsymbol{\sigma}_F$ given by Eq.  \eqref{eq:StressTotAndThermal} can be defined by the following lines.
\begin{lstlisting}
function $\sigma$_thermMod($\varepsilon$_in,s_in,T_in)
    $\varepsilon$Elas_in = $\varepsilon$_in - $\alpha$*(T_in-T0)*I2
   $\varepsilon$ElasTher = $\alpha$*(T0)*I2
    if tr($\varepsilon$Elas_in)  >= 0
      $\sigma$F = (s_in^2 + $\eta$)*$\sigma$_elas($\varepsilon$ElasTher)
  elseif tr($\varepsilon$Elas_in) < 0
      $\sigma$F = (s_in^2 + $\eta$)*P_dev $\odot$ $\sigma$_elas($\varepsilon$ElasTher) + P_vol$\odot$ $\sigma$_elas($\varepsilon$ElasTher) 
  end  
    return $\sigma$F
end
\end{lstlisting}
One can determine the positive part of the elastic free energy $\psi_{\text{elas}}^+ (\boldsymbol{\epsilon}^{\text{elas}})$ given by Eq.  \eqref{eq:HistoryFun} by writing the following function in Julia.
\begin{lstlisting}
function $\Psi$Pos($\varepsilon$_in,T_in)
    $\varepsilon$Elas_in = $\varepsilon$_in - $\alpha$*(T_in-T0)*I2
 if tr($\varepsilon$Elas_in)  >= 0
      $\Psi$Plus = 0.5*(($\varepsilon$Elas_in) $\odot$ $\sigma$_elas($\varepsilon$Elas_in))             
  elseif tr($\varepsilon$Elas_in)  < 0
      $\Psi$Plus = 0.5*((P_dev $\odot$ $\sigma$_elas($\varepsilon$Elas_in)) $\odot$ (P_dev $\odot$ ($\varepsilon$Elas_in))) 
    end
    return $\Psi$Plus
end
\end{lstlisting}
One needs to generate a discrete approximation of the finite element test and trial spaces of the problem on the discretized computational domain. Approximation of the finite element spaces associated with the phase field variable can be done by the following lines in Julia.
\begin{lstlisting}
order = 1
reffe_PF = ReferenceFE(lagrangian,Float64,order)
V0_PF = TestFESpace(model,reffe_PF;
  conformity=:H1)
U_PF = TrialFESpace(V0_PF)
sh = zero(V0_PF)
\end{lstlisting}
One can generate the approximation of the finite element spaces associated with the displacement field by writing the following lines in Julia.
\begin{lstlisting}
reffe_Disp = ReferenceFE(lagrangian,VectorValue{2,Float64},order)
        V0_Disp = TestFESpace(model,reffe_Disp;
          conformity=:H1,
          dirichlet_tags=["BottomEdge","TopEdge"],
          dirichlet_masks=[(true,true),(false,true)])
uh = zero(V0_Disp)
\end{lstlisting}
Similarly, finite element spaces for the temperature field can be generated as given below.
\begin{lstlisting}
reffe_Temp = ReferenceFE(lagrangian,Float64,order)
V0_Temp = TestFESpace(model,reffe_Temp;
  conformity=:H1,
  dirichlet_tags=["BottomEdge","TopEdge"])
\end{lstlisting}
We have defined a combined finite element space for displacement and temperature field to develop a multi-field solution algorithm using the following code. 
\begin{lstlisting}
V0 = MultiFieldFESpace([V0_Disp,V0_Temp])
\end{lstlisting}
To compute the integrals in the weak form given by Eq.  \eqref{eq:WeakFormParts} numerically, one needs to define an integration mesh along with an integration rule (for example, Gauss quadrature) in each of the cells in the triangulation. Using Gridap, one can easily define the integration mesh and the corresponding Lebesgue measure by using the built-in functions ``Triangulation" and ``Measure", respectively. For instance, one can use the following lines for computing the integrals in the weak form given by Eq.  \eqref{eq:WeakFormParts} defined in the domain $\Omega$ using a quadrature rule of degree two times the order of interpolation in the cells of the triangulation.
\begin{lstlisting}
degree = 2*order
$\Omega$ = Triangulation(model)
d$\Omega$ = Measure($\Omega$,degree)
\end{lstlisting}
One can find normal to a specific load boundary and define a measure on that boundary to compute a boundary integral using the following built-in functions available in Gridap.
\begin{lstlisting}
labels = get_face_labeling(model)
LoadTagId = get_tag_from_name(labels,"TopEdge")
$\Gamma$_Load = BoundaryTriangulation(model,tags = LoadTagId)
d$\Gamma$_Load = Measure($\Gamma$_Load,degree)
n_$\Gamma$_Load = get_normal_vector($\Gamma$_Load)
\end{lstlisting}
To find the values of variables that are defined by using the built-in function ``CellState" at the Gauss points, one need to use the ``project" function as defined below.
\begin{lstlisting}
function project($q$,model,d$\Omega$,order)
  reffe = ReferenceFE(lagrangian,Float64,order)
  $\text{V}$ = FESpace(model,reffe,conformity=:L2)
  a($u$,$v$) = $\int$($u$*$v$)*d$\Omega$
  b($v$) = $\int$($v$*$q$)*d$\Omega$
  op = AffineFEOperator(a,b,$\text{V}$,$\text{V}$)
  qh = Gridap.solve(op)
  return qh
end
\end{lstlisting}
In the present study, a staggered scheme is used to update the solution from the pseudo time $t_n$ to $t_{n+1}$ \cite{miehe2010phase}.       Given the displacement and temperature field and the history function at the time $t_n$, one can update the phase-field at the time $t_{n+1}$ by using the following function.
\begin{lstlisting}
function  stepPhaseField(uh_in,$\Psi$PlusPrev_in)
        a_PF(s,$\Phi$) = $\int$( Gc*lsp*$\nabla$($\Phi$)$\cdot$ $\nabla$(s) + 2*$\Psi$PlusPrev_in*s*$\Phi$  + (Gc/lsp)*s*$\Phi$ )*d$\Omega$
        b_PF($\Phi$) = $\int$( (Gc/lsp)*$\Phi$ )*d$\Omega$
        op_PF = AffineFEOperator(a_PF,b_PF,U_PF,V0_PF)
        sh_out = Gridap.solve(op_PF)           
    return sh_out
    end
\end{lstlisting}
Following function has been used for defining the product between the conductivity tensor and gradient of temperature in the weak form corresponding to temperature field (see Eq.  \eqref{eq:WeakForm}).
\begin{lstlisting}
$\kappa$GradTemp($\nabla$,s_in) = (s_in^2 + $\eta$)*$\kappa$_mat*$\nabla$
\end{lstlisting}
Approximating the time derivative of temperature by using backward difference method as 
\begin{equation}
\dot{T}  = \frac{T_{n+1}-T_n}{t_{n+1}-t_n}
   = \frac{T_{n+1}-T_n}{\Delta t},
 \end{equation}
one can update the displacement vector and temperature field at time $t_{n+1}$ by solving coupled set of equations using the values of temperature vector and the history function at time $t_n$, and the computed value of phase-field at time $t_{n+1}$. Using a built-in function MultiFieldFESpace in Gridap, we have written a function for updating the displacement and temperature as given below. 
\begin{lstlisting}
  function   stepDispTemp(uh_in,sh_in,T_in,vApp,TApp,delt)
        uApp1(x) = VectorValue(0.0,0.0)
        uApp2(x) = VectorValue(0.0,vApp)
        U_Disp = TrialFESpace(V0_Disp,[uApp1,uApp2])
        Tapp1(x) = T0
        Tapp2(x) = TApp 
        Tg = TrialFESpace(V0_Temp,[Tapp1, Tapp2])
        U = MultiFieldFESpace([U_Disp,Tg])
        a((u,T),(v,w)) = $\int$( ($\varepsilon$(v) $\odot$ ($\sigma$_totMod$\circ$($\varepsilon$(u),$\varepsilon$(uh_in),sh_in,T,T_in))) + $\nabla$(w)$\cdot$($\kappa$GradTemp$\circ$ ($\nabla$(T),sh_in)) + (($\rho$*c*T*w)/delt))*d$\Omega$
        b((v,w)) = $\int$((($\rho$*c*T_in*w)/delt) - ($\varepsilon$(v) $\odot$ ($\sigma$_thermMod$\circ$($\varepsilon$(uh_in),sh_in,T_in))))*d$\Omega$
        op = AffineFEOperator(a,b,U,V0)
        uhTh = Gridap.solve(op)                
        uh_out,Th_out =  uhTh
    
    return uh_out,Th_out
end
\end{lstlisting}
Once both displacement and phase-field are determined at time $t_{n+1}$, one can update the energy history function at time $t_{n+1}$ by defining it as
\begin{equation}
 \mathcal{H}_{n+1} = 
 \begin{cases}
    \mathcal{E}_{n+1}\,\,\text{for}\,\, \mathcal{E}_{n+1} > \mathcal{E}_{n}& \\
    \mathcal{E}_{n} \,\,\text{otherwise},
  \end{cases}
 \end{equation}
where the history function $\mathcal{H}(\mathcal{E})$ of elastic free energy $\mathcal{E} =\psi_{\text{elas}}^+ (\boldsymbol{\epsilon}^{\text{elas}})$ need to be defined in Julia (see Section \ref{sec:BVPdef}) that can be achieved by writing the following lines.
\begin{lstlisting}
function new_EnergyState($\psi$PlusPrev_in,$\psi$hPos_in)
  $\psi$Plus_in = $\psi$hPos_in
  if $\psi$Plus_in >= $\psi$PlusPrev_in
    $\psi$Plus_out = $\psi$Plus_in
  else
    $\psi$Plus_out = $\psi$PlusPrev_in
  end
  true,$\psi$Plus_out
end
\end{lstlisting}
Finally, one can simulate the thermo-mechanical brittle fracture in the notched plate by applying a monotonic displacement control loading and solve for the unknown field variables at each loading step by writing the main routine in Julia that uses the above-defined functions as listed below.  
\begin{lstlisting}[language=C,backgroundcolor=\color{gray!5!white},showstringspaces=false,breaklines=true,basicstyle=\ttfamily]
t = 0.0
innerMax = 10
count = 0
tol = 1e-8
Load = Float64[]
Displacement = Float64[]
push!(Load, 0.0)
push!(Displacement, 0.0)
$\Psi$PlusPrev = CellState(0.0,d$\Omega$) 
sPrev = CellState(1.0,d$\Omega$)
sh = project(sPrev,model,d$\Omega$,order)
ThPrev = CellState(T0,d$\Omega$)
Th = project(ThPrev,model,d$\Omega$,order)
while t .< tMax 
    count = count .+ 1      
    t = t + delt
    vApp = AppVel*t    
    TApp = TAppVec[count]
 for inner = 1:innerMax   
$\Psi$hPlusPrev = project($\Psi$PlusPrev,model,d$\Omega$,order)
        RelErr = abs(sum($\int$( Gc*lsp*$\nabla$(sh)$\cdot$ $\nabla$(sh) + 2*$\Psi$hPlusPrev*sh*sh  + (Gc/lsp)*sh*sh)*d$\Omega$ - $\int$( (Gc/lsp)*sh)*d$\Omega$))/abs(sum($\int$( (Gc/lsp)*sh)*d$\Omega$))
        sh = stepPhaseField(uh,$\Psi$hPlusPrev) 
        uh,Th = stepDispTemp(uh,sh,Th,vApp,TApp,delt)

        $\Psi$hPos_in = $\Psi$Pos$\circ$($\varepsilon$(uh),Th)   
        
        update_state!(new_EnergyState,$\Psi$PlusPrev,$\Psi$hPos_in)
  
        if RelErr < tol
            break 
        end      
    end
    
  Node_Force = sum($\int$( n_$\Gamma$_Load $\cdot$ ($\sigma$_elasMod$\circ$($\varepsilon$(uh),$\varepsilon$(uh),sh,Th,Th)) ) *d$\Gamma$_Load) 
    
    push!(Load, Node_Force[2])     
    push!(Displacement, vApp)           

end 
\end{lstlisting}
One can create output files at each loading step to view the results in ParaView. For instance, one can write a ``.vtu" file to save data for the solution of the displacement field, phase-field, and temperature field at each loading step by using the following lines in Julia. 
\begin{lstlisting}
writevtk($\Omega$,"results_PhaseFieldThermoElastic",cellfields=
["uh"=>uh,"s"=>sh,"T"=>Th])
\end{lstlisting}
One may generate the load-displacement curve by using the plot command as given below.
\begin{lstlisting}
plot(Displacement,Load)
\end{lstlisting}

\begin{figure}[h]
    \centering
    \includegraphics[scale = 2.5]{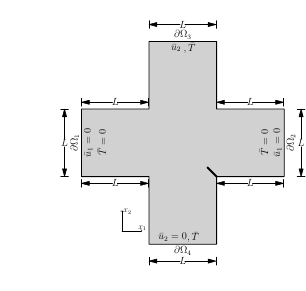}
    \caption{Geometry and boundary conditions for a cruciform shape notched plate ($\bar{u}_1$ and $\bar{u}_2$ corresponds to the specified displacement along the direction $x_1$ and $x_2$ , respectively, $\bar{T}$ corresponds to the specified temperature, $\partial \Omega_1$,\,$\partial \Omega_2$,\,$\partial \Omega_3$,\,$\partial \Omega_4$ denotes left, right, top, and bottom boundaries, respectively).}
    \label{fig:CruciformGeom}
\end{figure}
\section{Numerical simulations}
\label{Sec:NumericalSimulation}
In this section, we have considered a few standard problems available in the literature for validating our proposed phase-field model and its Julia implementation. The validation problems include cruciform shaped material subjected to different boundary conditions, single edge notched specimen under tension, bi-material beam subjected to reduction of temperature, and a ceramic beam quenched to lower temperature. We have used constant strain triangle (CST) elements for finite element solutions of the problems. The simulation results show the capability of the proposed model to reproduce the results available in the literature.

\subsection{Cruciform shape plate with notch}
\label{sec:cruciform}
We have considered a plate of cruciform shape with a notch as shown in Fig. \ref{fig:CruciformGeom}. The problem has been introduced by Prasad {\it{et al.}} \cite{prasad1994incremental}, and since then, it has been considered as a benchmark problem for numerical analysis of thermo-mechanical fracture. We have taken $L$ as $50$ mm and chosen material properties from  \cite{prasad1994incremental} and \cite{mandal2021fracture}, which are $\emph{E} = 218400$ $N/m^2$, $\nu = 0.2$, $\emph{G}_c = 0.0002\, N/m$, $\rho = 0.0$, $\alpha = 6\times10^{-4}\,\tccentigrade$, $k_0 = 1.0\,W/m\tccentigrade$,  $c = 1.0 J/kg\, \tccentigrade$ . For numerical simulations, we have assumed plane stress condition similar to Mandal {\it{et al.}} \cite{mandal2021fracture}, and we have discretized the domain using CST elements. The Julia code for creating the mesh file for a particular case is given in \href{https://github.com/MdMasiurRahaman/PhaseFieldModelForThermoMechanicalFracture-/blob/main/MeshCrushiformShapeWithInclinedCrack.ipynb}{Julia code for meshing of cruciform shape notched plate}. 

We have considered the following three loading conditions, including the boundary conditions at the edges for all three cases shown in Fig. \ref{fig:CruciformGeom}.
\begin{enumerate}[a.]
    \item  Application of quasi-static load in the form of incremental displacement of $\bar{u}_2=3.5\times10^{-7}$m at each time step on $\partial \Omega_3$, with $\bar{T} = 0$ on $\partial \Omega_3$ and $\partial \Omega_4$ for the consideration of pure mechanical loading case.
    \item Increasing the temperature on $\partial \Omega_3$ with simultaneous reduction on $\partial \Omega_4$ without any restriction on $\bar{u}_2$ on $\partial \Omega_3$  to consider a pure thermal loading case.  At each time step, temperature increases by $0.1 \tccentigrade$ on $\partial \Omega_3$  while reducing the temperature on $\partial \Omega_4$  by $-0.1 \tccentigrade$ and keeping the temperature on boundaries $\partial \Omega_1$  and $\partial \Omega_2$  at the reference temperature.
    \item The combined effect of mechanical and thermal loading cases. With increment in displacement as $\bar{u}_2 = 1.92\times10^{-7}$m on $\partial \Omega_3$  and temperature increment of $0.06 \tccentigrade$ and $-0.06 \tccentigrade$ on $\partial \Omega_3$  and $\partial \Omega_4$ , respectively at each time step. 
\end{enumerate}

\begin{figure}[ht!]
  \centering
  \subfloat[]{
	\begin{minipage}[c][0.99\width]{0.35\textwidth}
	 \includegraphics[width=\textwidth]{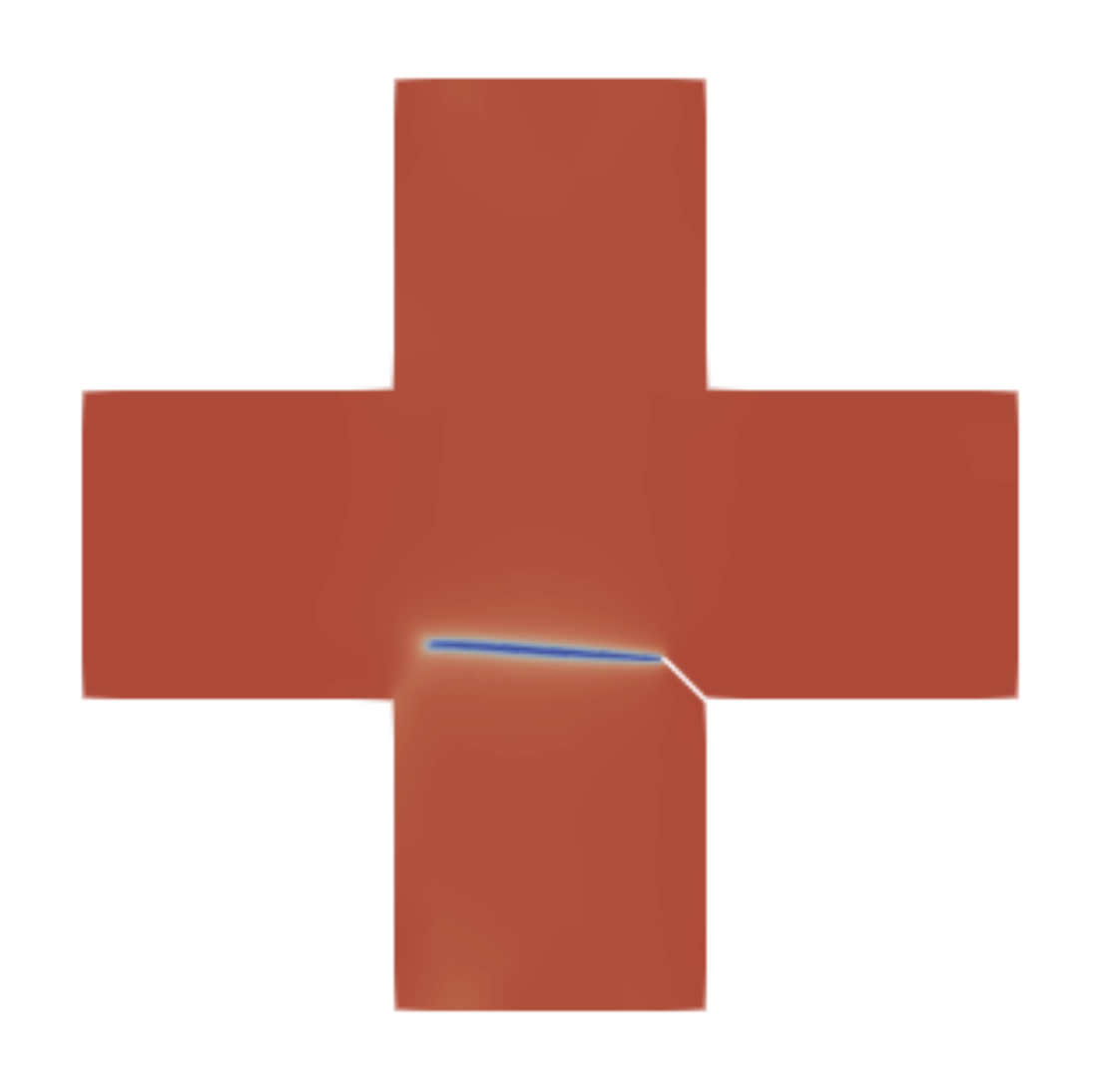}
	\end{minipage}}
		\newline
	\newline
		\newline
	  \centering
  \subfloat[]{
	\begin{minipage}[c][0.99\width]{0.35\textwidth}
	 \includegraphics[width=\textwidth]{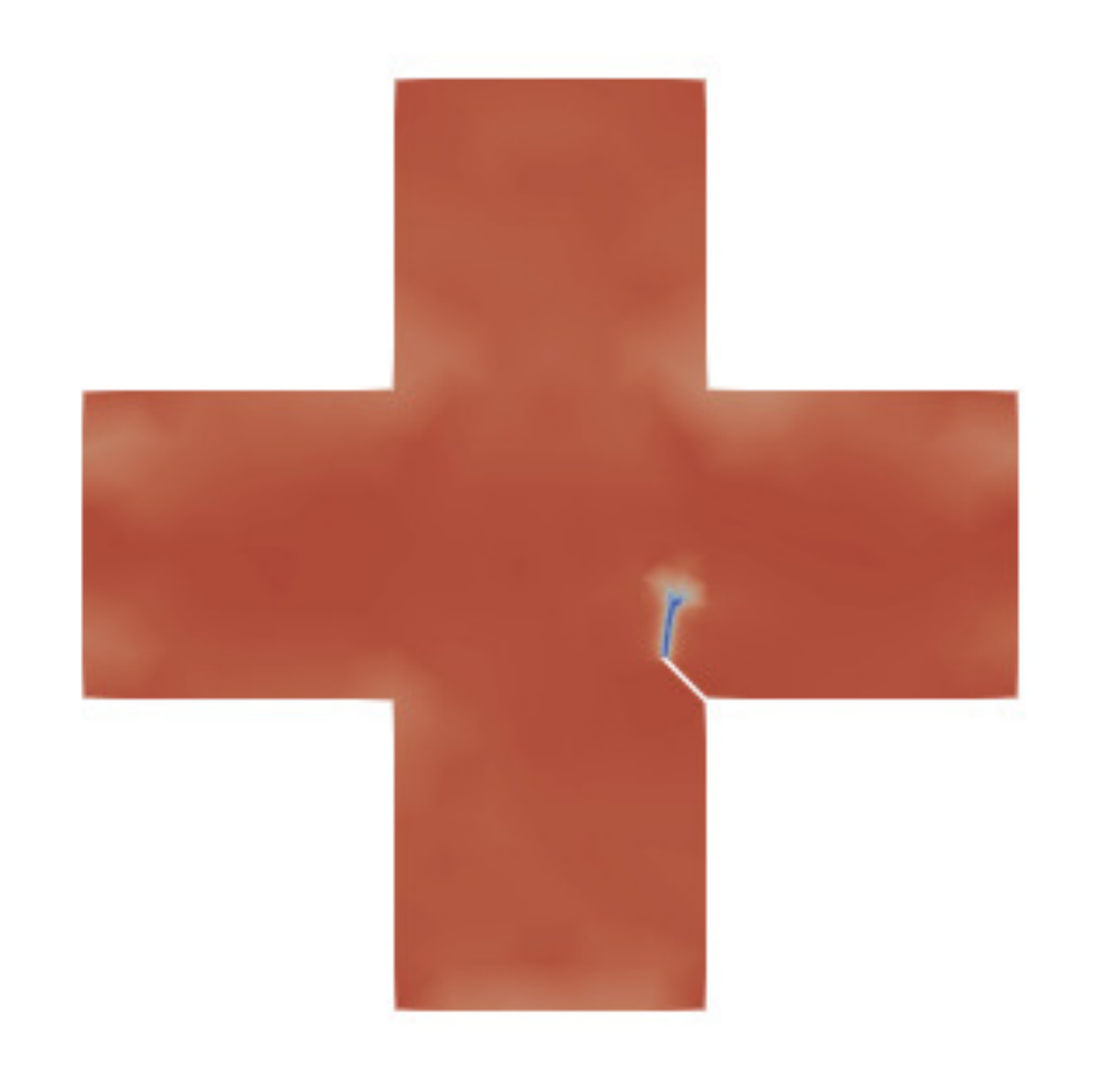}
	\end{minipage}}
	\newline
	\newline
		\newline
	  \centering
	\subfloat[]{
	\begin{minipage}[c][0.99\width]{0.35\textwidth}
	 \includegraphics[width=\textwidth]{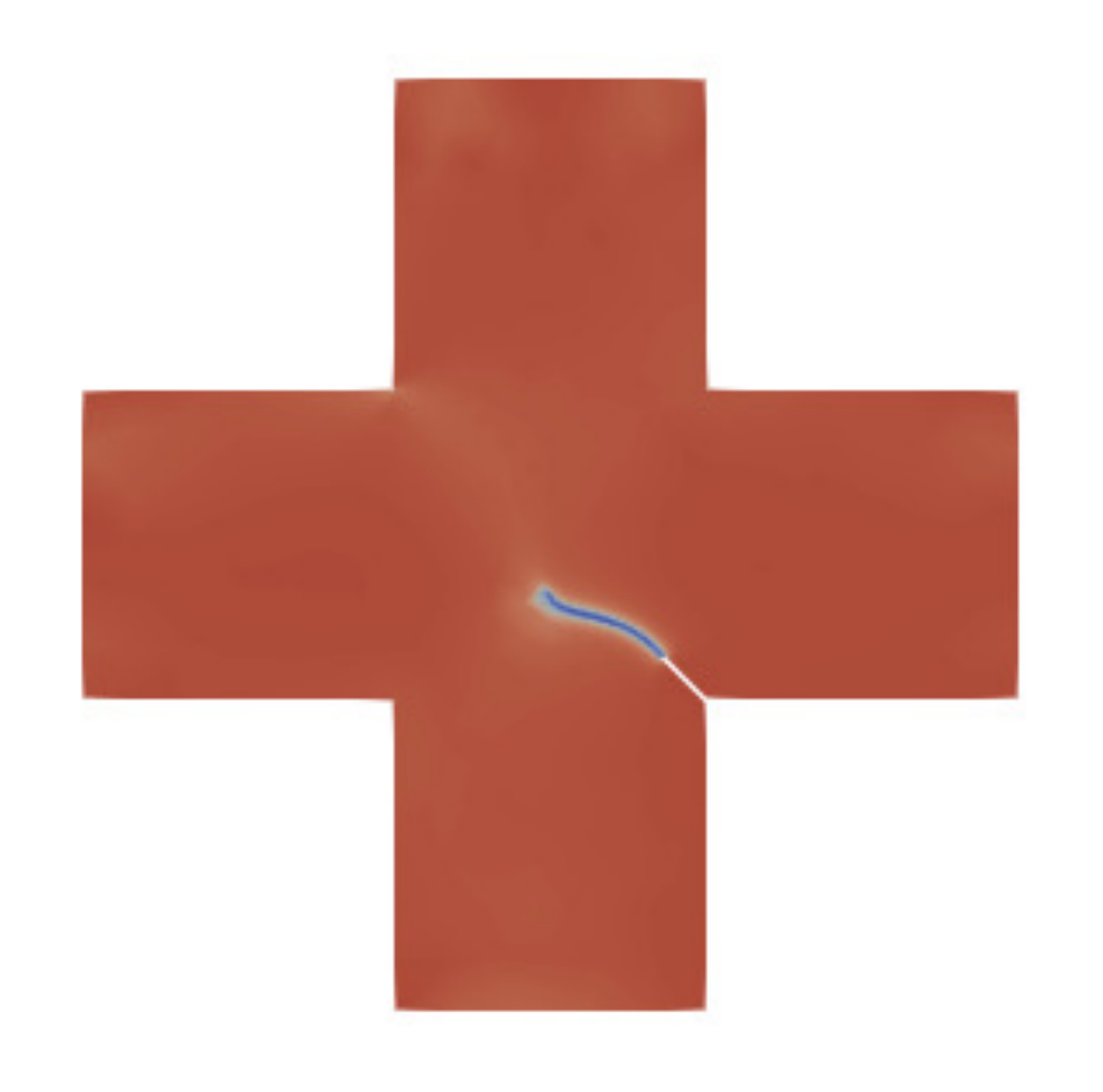}
	\end{minipage}}
\caption{Simulated fracture results for cruciform specimen for (a) pure mechanical loading ($3.4\times10^{-5}$m)
(b) pure thermal loading at ($8\tccentigrade$) and (c) combined mechanical and thermal loading ($1.53\times10^{-5}m\,\, \text{and}\,\, 4.8\tccentigrade$)}
\label{fig:CrusiformSpecimen}
\end{figure}
The length of the notch was taken to be $0.2L$ with an inclination of $135^{\circ}$ counterclockwise from horizontal. The length scale parameter $l_s$ was taken to be $5\times10^{-4}$m.  Furthermore, a refined mesh was used in the expected portion of the domain where the crack may propagate with the edge length of $l_s/4$. The obtained crack pattern for all the three cases is demonstrated in Fig. \ref{fig:CrusiformSpecimen}. As can be seen from Fig. \ref{fig:CrusiformSpecimen}, the crack propagation for pure mechanical loading case is completely horizontal, where for the thermal loading case the crack initiates vertically from the notch and eventually starts bending right. For the combined case the crack propagates in between the purely mechanical and pure thermal case i.e. the crack propagates diagonally. The obtained crack propagation directions shows similar trend to that obtained by Duflot \cite{duflot2008extended}, Prasad {\it{et al.}} \cite{prasad1994incremental}, and Mandal {\it{et al.}} \cite{mandal2021fracture}.

\subsection{Single edge notched plate under tension}
\label{sec:SENT}
\begin{figure}[ht!]
 \centering
 \includegraphics[width=0.42\textwidth]{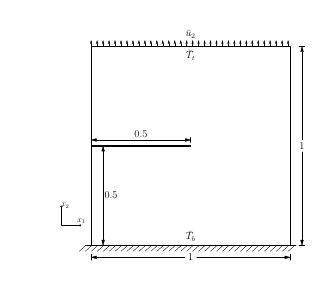}
    \caption{Geometry and boundary conditions for a single edge notched plate under tension ($\bar{T}_t$ and $\bar{T}_b$ are the specified temperature at the top and bottom boundary, respectively, $\bar{\boldsymbol{u}}_2$ is the specified displacement along the direction $x_2$, dimensions are in mm).}
\label{fig:SENTGeom}
\end{figure}

We have considered a notched plate as shown in Fig. \ref{fig:SENTGeom} and chosen the material properties as  $\emph{E} = 32\times10^9$ $N/m^2$, $\nu = 0.2$, $\emph{G}_c = 3\, N/m$, $\rho = 2450 kg/m^3$, $\alpha = 8\times10^{-6}\,\tccentigrade$, $k_0 = 300\,W/m\tccentigrade$,  $c = 0.775 J/kg\, \tccentigrade$ , taken from \cite{tangella2021hybrid}.  We have used CST elements for the finite element discretization of the notched plate and the Julia code for creating the mesh file is given in \href{https://github.com/MdMasiurRahaman/PhaseFieldModelForThermoMechanicalFracture-/blob/main/MeshNotchedPlateUnderTension.ipynb}{Julia code for meshing of single edge notched plate}. 

\begin{figure}[t]
  \subfloat[]{
	\begin{minipage}[c][0.9\width]{0.37\textwidth}
	   \includegraphics[width=\textwidth]{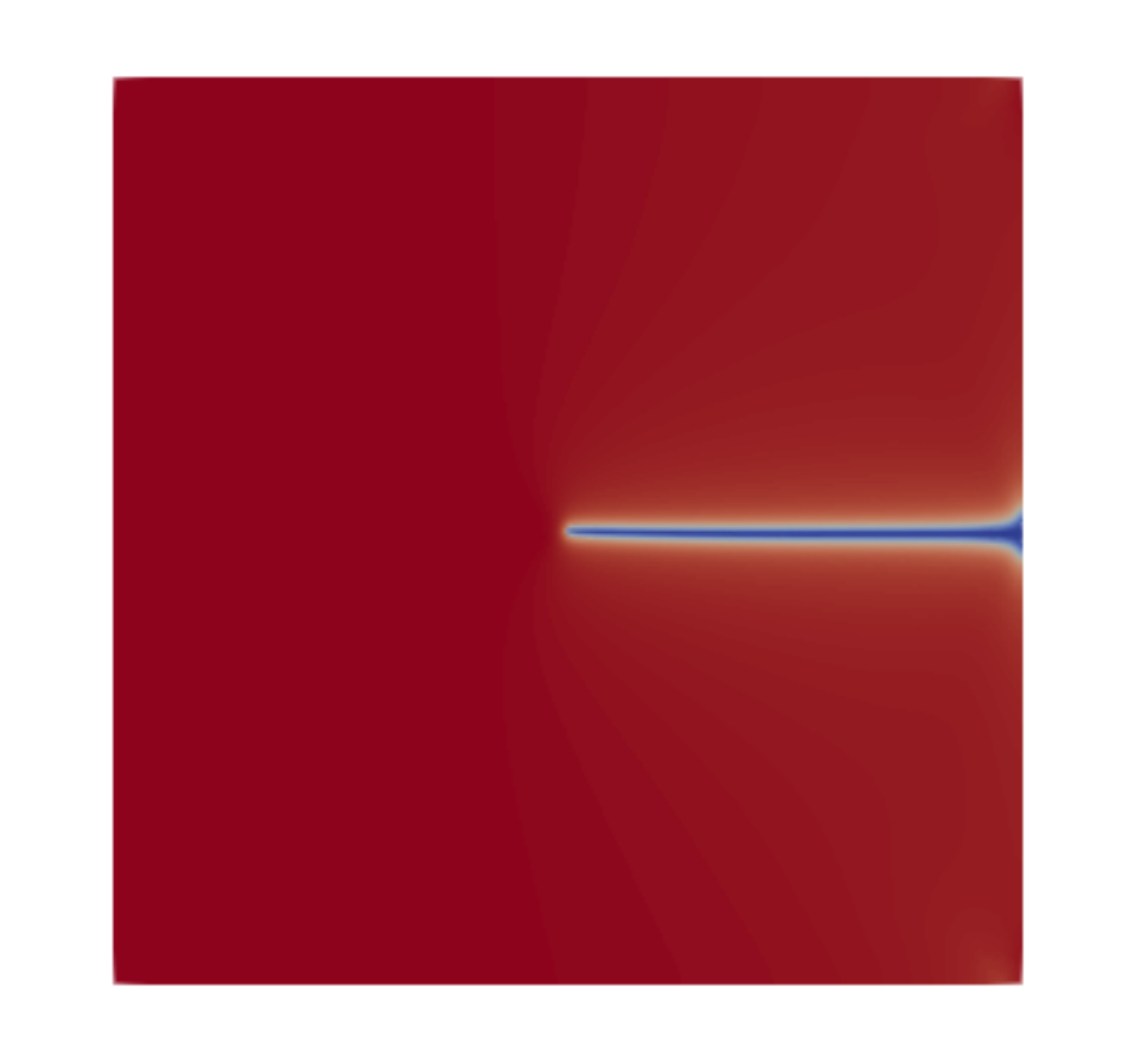}
	\end{minipage}}
	\newline
  \subfloat[]{
	\begin{minipage}[c][0.9\width]{0.37\textwidth}
	 \includegraphics[width=\textwidth]{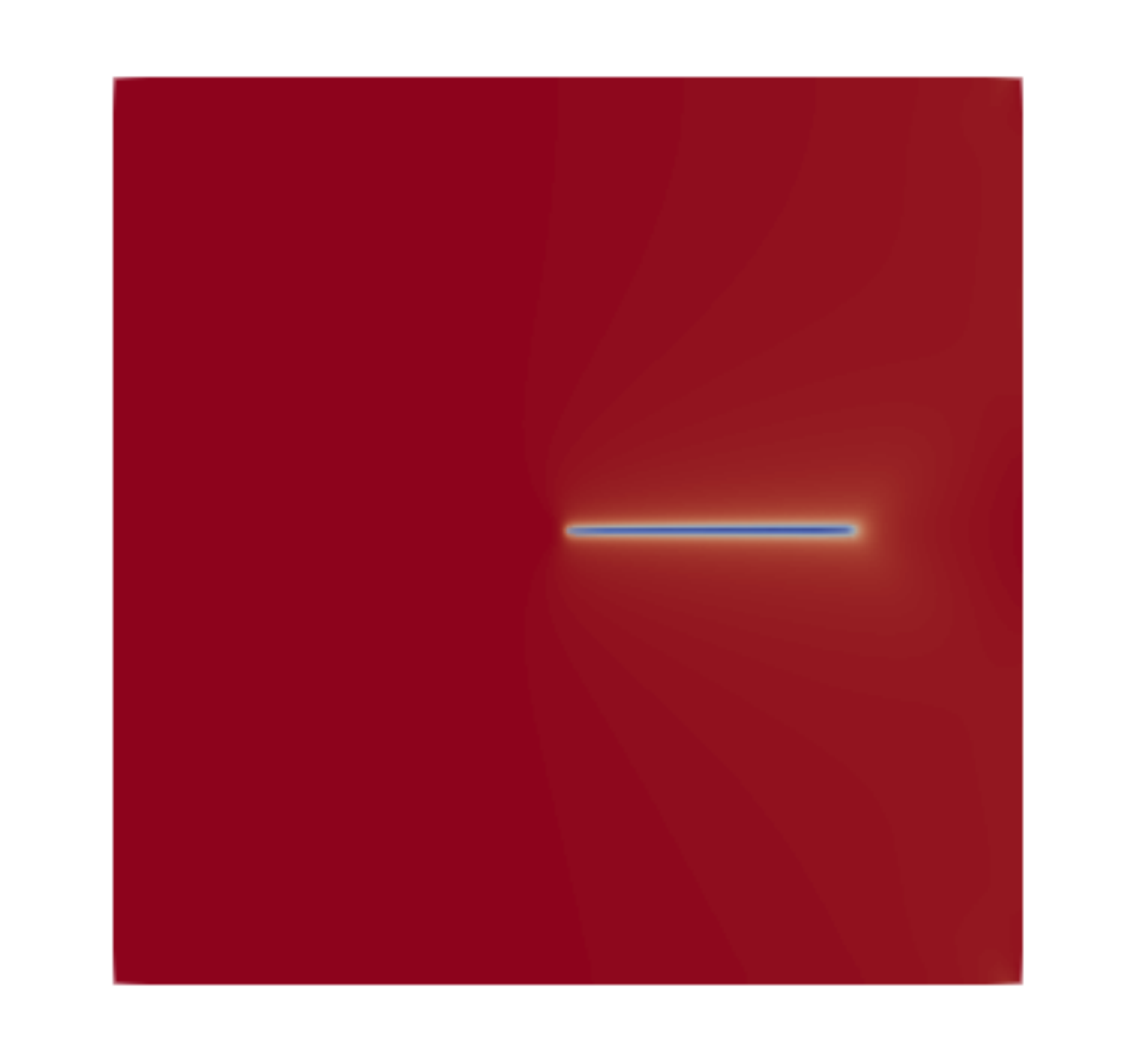}
	\end{minipage}}
		\newline
	\subfloat[]{
	\begin{minipage}[c][0.9\width]{0.37\textwidth}
	 \includegraphics[width=\textwidth]{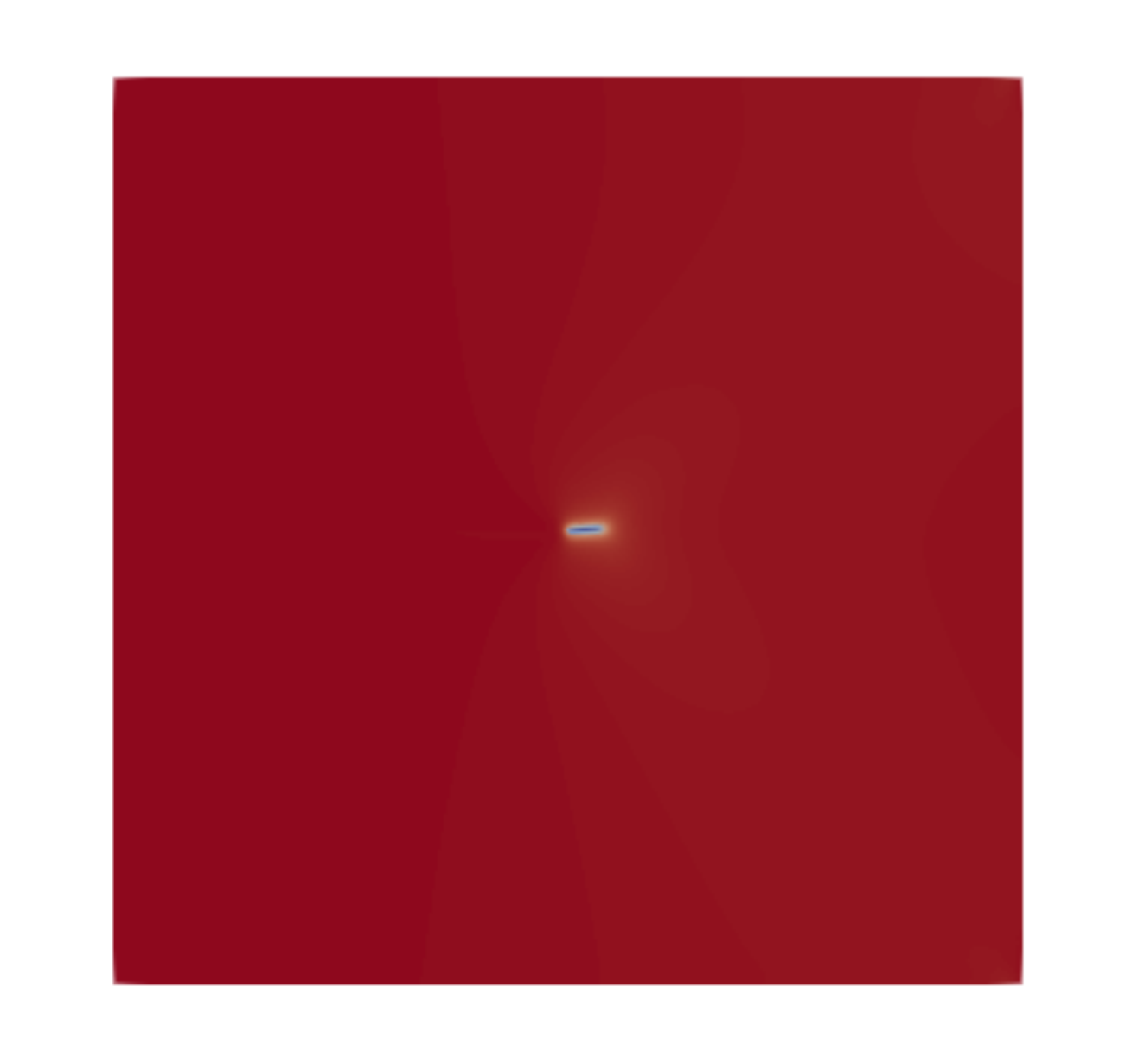}
	\end{minipage}}
\caption{Simulated fracture results for single edge notched plate under tension at an applied displacement $\bar{\boldsymbol{u}}_2 = 8.4\times10^{-4} mm.$ and applied temperature increment (a) $\Delta T = -75\tccentigrade$ (b) $\Delta T = 0\tccentigrade$  and (c) $\Delta T = +75\tccentigrade$.}
\label{fig:SENPCrack}
\end{figure}
We have reproduced the results for the following cases assuming plane strain condition \cite{tangella2021hybrid}:
\begin{enumerate}[a.]
    \item  Application of quasi-static load in the form of incremental displacement at the top boundary, with restrained displacement boundary conditions (see Fig.  \ref{fig:SENTGeom}) while maintaining the temperature of the domain at $300\,\tccentigrade$ for the consideration of pure mechanical loading case.
    \item Incrementing $\bar{T}_t$ upto $(T_0+50)\tccentigrade$ and keeping $\bar{T}_b$ at $T_0$  for the consideration of thermo-mechanical coupling along with incremental displacement at top. A detailed description of the implementation of this case has been discussed in section \ref{sec:JuliaImplementation}. 
    \label{it:sentA}
    \item Decrementing $\bar{T}_t$ upto $(T_0-50)\tccentigrade$ and keeping $\bar{T}_b$ at $T_0$  for the consideration of thermo-mechnical coupling under opposite gradient along with incremental displacement at top.
    \label{it:sentB}
    \item Incrementing $\bar{T}_t$ up to $(T_0+75)\tccentigrade$ and keeping $\bar{T}_b$ at $T_0$  for the consideration of thermo-mechanical coupling with higher temperature gradient than the previous (see \ref{it:sentA}) along with incremental displacement at top.
    \item Decrementing $\bar{T}_t$ upto $(T_0-75)\tccentigrade$ and keeping $\bar{T}_b$ at $T_0$  for the consideration of thermo-mechnical coupling under opposite higher temperature gradient than previous (see \ref{it:sentB}) gradient along with incremental displacement at top.    
    \end{enumerate}
\begin{figure}[ht]
    \centering
    \includegraphics[scale = 0.4]{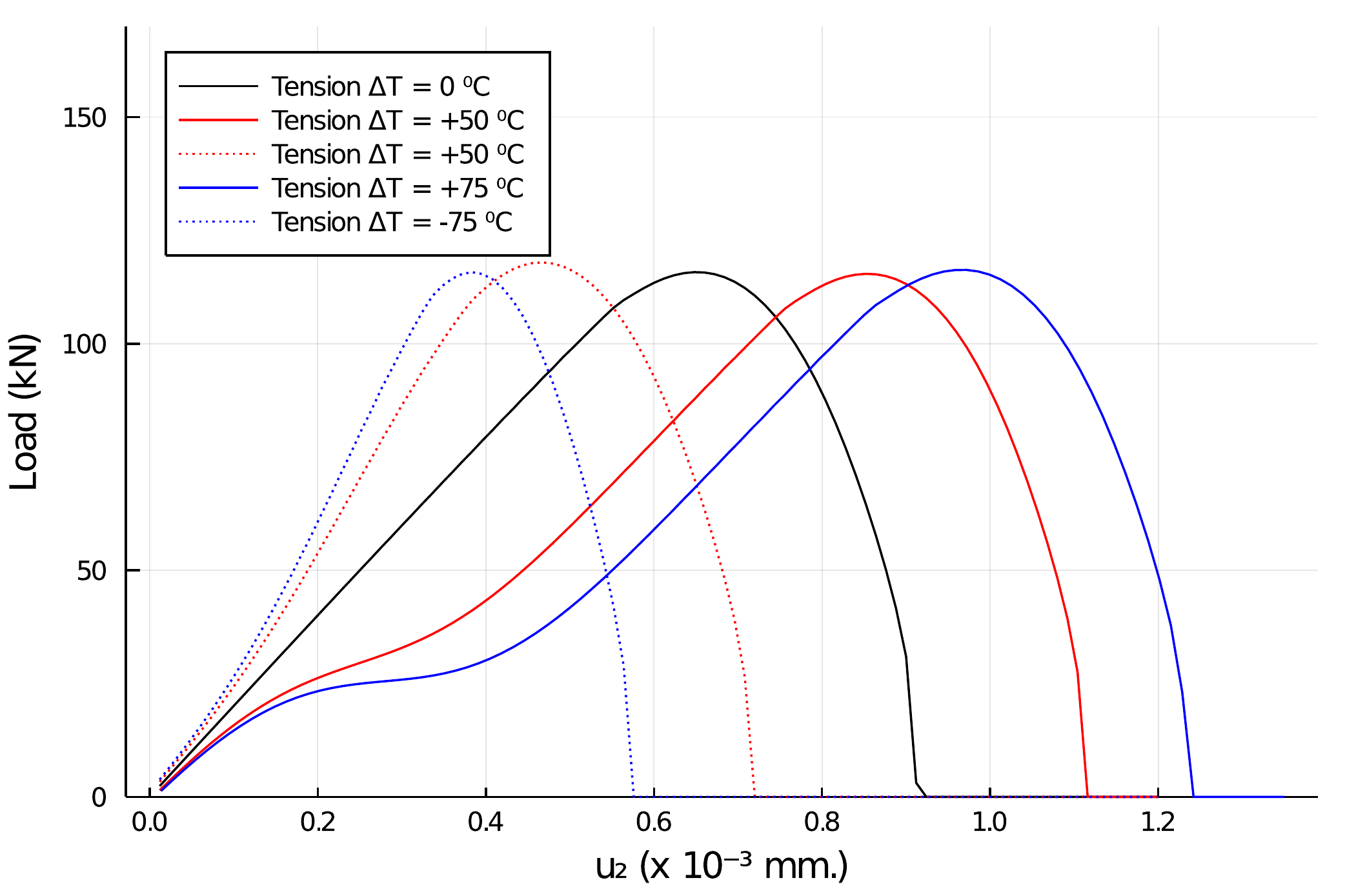}
    \caption{Load-displacement curve for single edge notched plate under tension for $\Delta T$ = $-75\tccentigrade$, $-50\tccentigrade$, $0\tccentigrade$, $+50\tccentigrade$, $+75\tccentigrade$.}
    \label{fig:ThermoMechanicalEffectonSENPunderTensionGraph}
\end{figure}
The crack propagation for load case (a), (d) and (e) has been shown in Fig. \ref{fig:SENPCrack}. The delay in crack propagation was observed with the increasing temperature at the top boundary, which can also be seen in the load-displacement plots with the delay in peak (see Fig. \ref{fig:ThermoMechanicalEffectonSENPunderTensionGraph}). Case (b) shows a similar behavior as case (d) by showing delay in crack propagation in comparison to pure mechanical case but apparent early crack propagation in comparison to case (d) characterized by the load peak in-between case (a) and case (d). And since in case (c) the temperature is in between $0 \tccentigrade$ and $-75\tccentigrade$ the position of load peak shows that the obtained crack position will also be in between the one obtained for case (a) and case (e).  We have reproduced the results reported in Tangella {\it{et al.}} \cite{tangella2021hybrid}.

\subsection{Notched bi-material beam}
\begin{figure}[ht!]
    \begin{subfigure}[ht!]{0.5\textwidth}
    \centering
 \includegraphics[width=9cm,height=2.5cm]{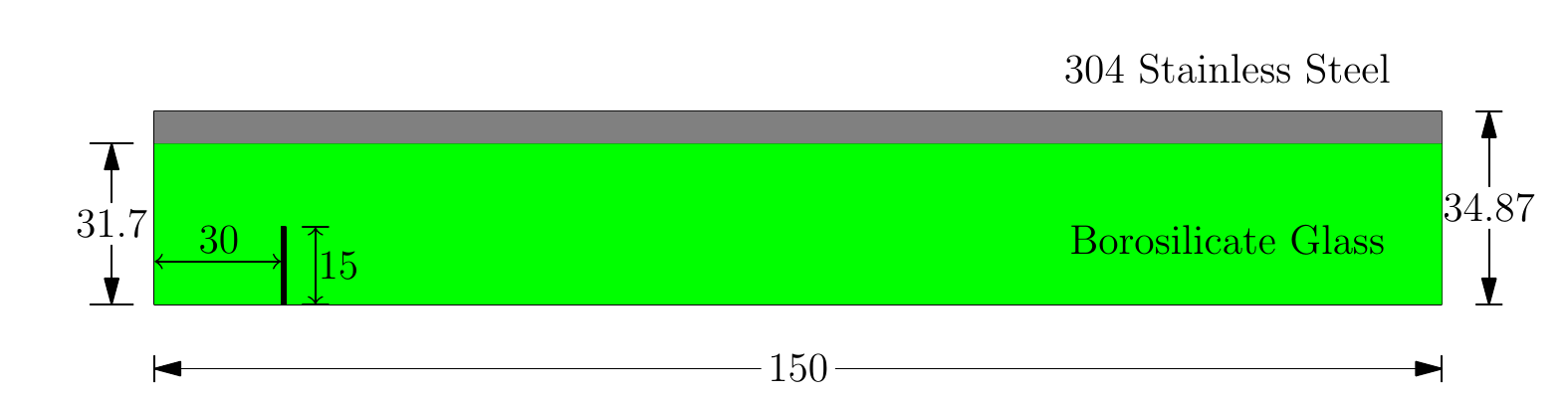}
         \caption{}
         \label{fig:BiMatGeom}
     \end{subfigure}
     \newline
    \begin{subfigure}[ht!]{0.5\textwidth}
    \centering
 \includegraphics[width=9cm,height=2.5cm]{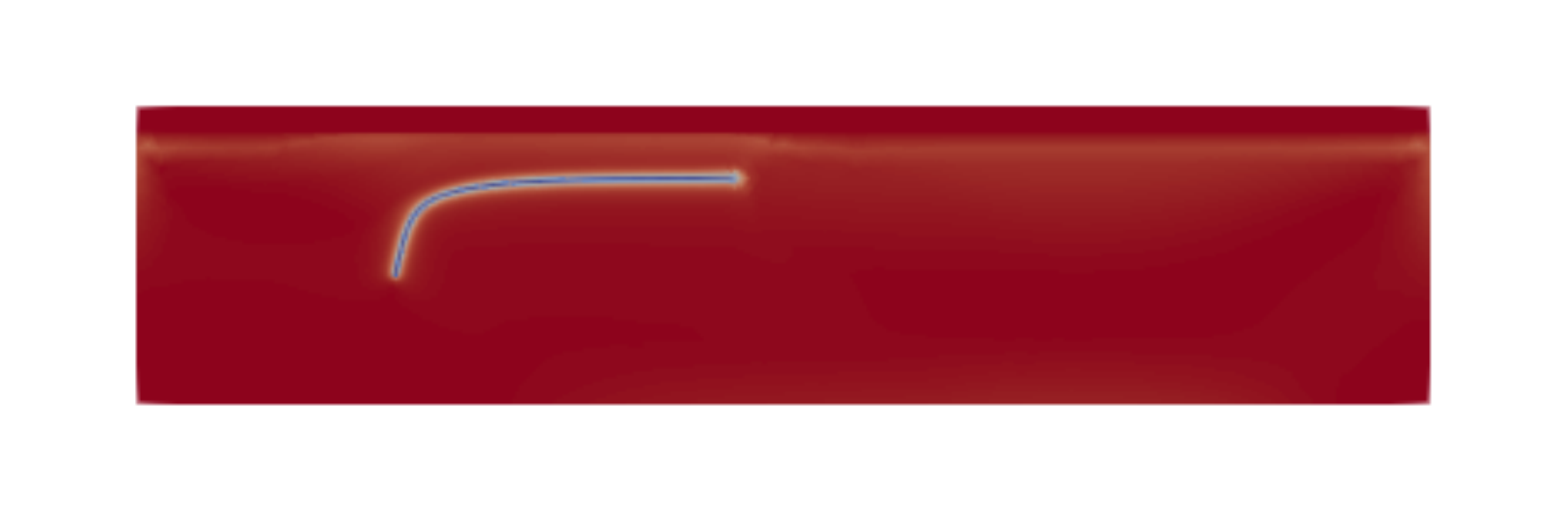}
         \caption{}
          \label{fig:BiMatCrack}
     \end{subfigure}
  \caption{(a) Geometry of bi-material beam (dimensions are in mm) (b) crack propagation in the bi-material beam due to gradual reduction of temperature inducing due to the difference in temperature coefficient of the materials.}
\label{fig:BimaterialGeometryandCrack}
\end{figure}

\begin{figure}[ht!]
 \centering
    \begin{subfigure}[b]{0.25\textwidth}
         \centering
 \includegraphics[width=0.9\textwidth]{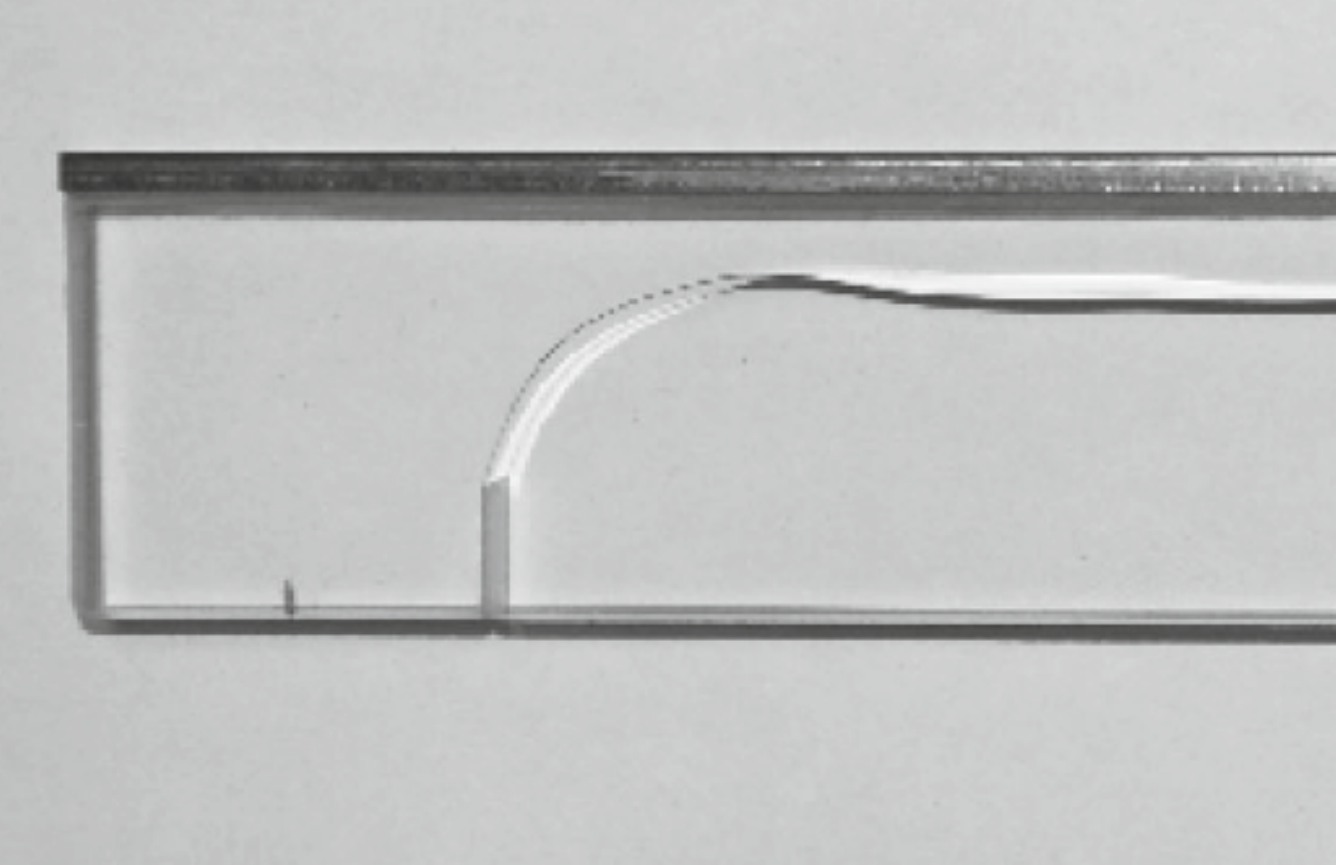}
         \caption{}
         \label{fig:BiMatExp}
     \end{subfigure}
     \begin{subfigure}[b]{0.275\textwidth}
         \centering
 \includegraphics[width=0.9\textwidth]{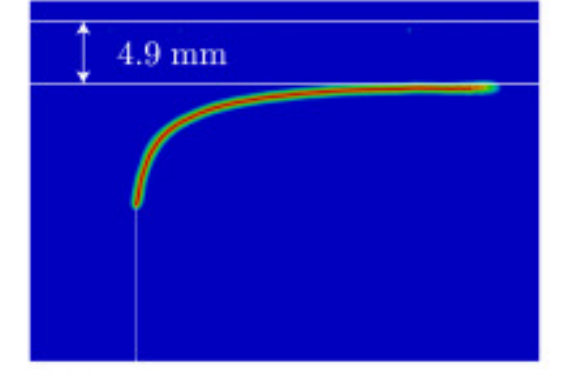}
         \caption{}
         \label{fig:BiMatNum}
     \end{subfigure}
    \begin{subfigure}[b]{0.25\textwidth}
         \centering
 \includegraphics[width=0.95\textwidth]{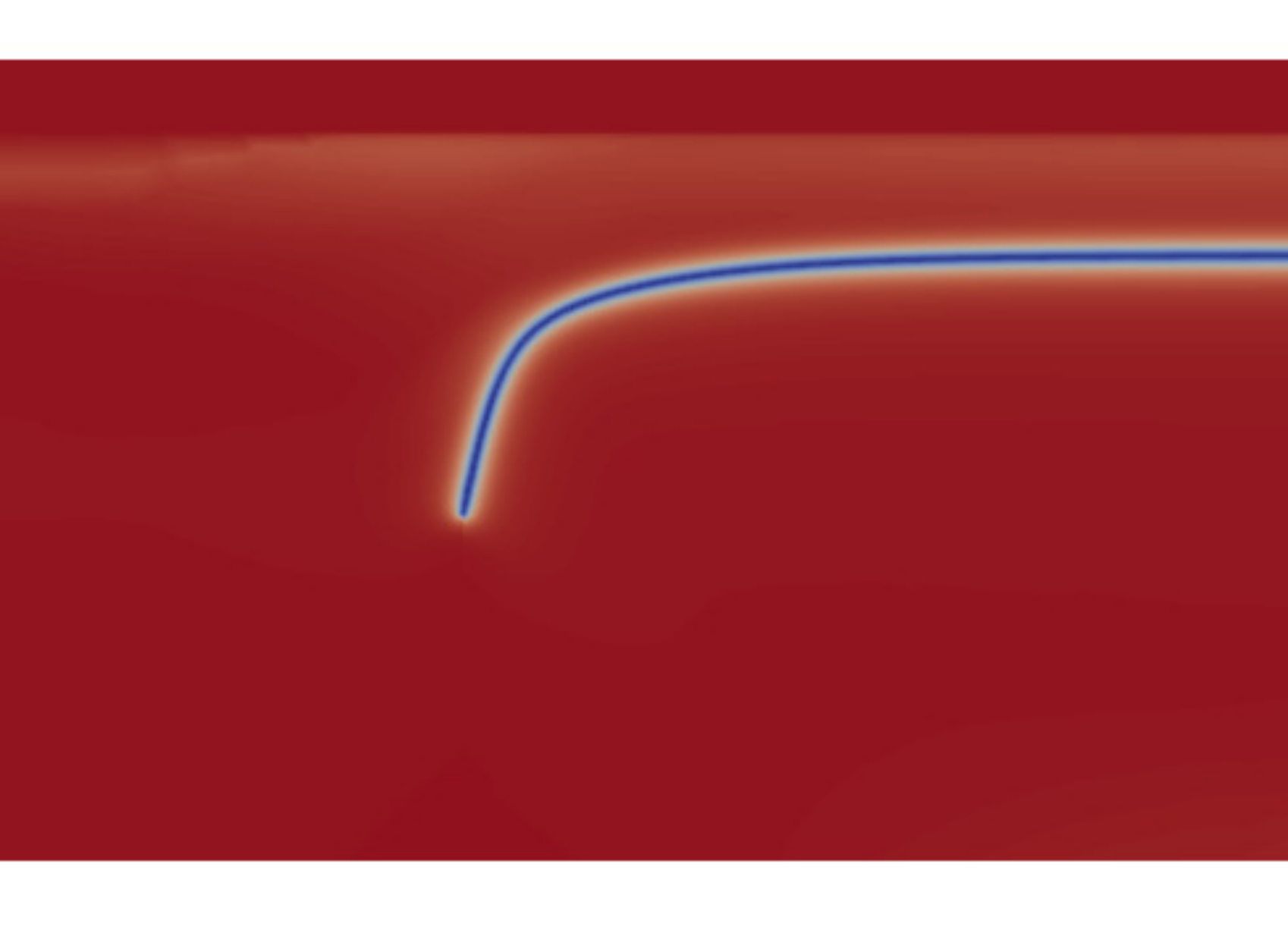}
         \caption{}
          \label{fig:BiMatTheor}
     \end{subfigure}
  \caption{Crack path in a bi-material beam: (a) experimentally observed \cite{grutzik2018crack} (b) prediction by the PF-CZM model \cite{mandal2021fracture} and (c) prediction using the proposed model.}
\label{fig:BimaterialComparison}
\end{figure}

We have chosen a bi-material beam as reported in \cite{grutzik2018crack} to simulate the behavior of fracture due to the dissimilar thermal coefficient of the materials and the advantage of using gridap in case of multi-material problems by defining the multi-material constitutive laws based on the assigned material tags. The geometry of the considered model has been shown in Fig. \ref{fig:BiMatGeom}. The crack position, height and material properties is chosen from Mandal {\it{et al.}} \cite{mandal2021fracture}, which are $\emph{E} = 64\times10^3$ MPa, $\nu = 0.2$, $\emph{G}_c = 0.4\, N/mm$, $\alpha = 3.25\times10^{-6}\,\tccentigrade$ corresponding to borosilicate glass, and $\emph{E} = 193\times10^3$ MPa, $\nu = 0.29$, $\alpha = 193\times10^{-6}\,\tccentigrade$ corresponding to $304$ stainless steel. We have considered the length of the specimen as $150$ mm. A notch of width $1$ mm and height $15$ mm has been assumed at a distance of $30$ mm from the left edge. The length scale parameter $l_s$ has been chosen to be $0.3$ mm; the mesh is refined in the area where the crack propagation was expected with the elements of size $l_s/4$. Plane stress condition has been assumed for the simulation. Julia code for the finite element meshing of the specimen is given in \href{https://github.com/MdMasiurRahaman/PhaseFieldModelForThermoMechanicalFracture-/blob/main/MeshBimaterialBeam.ipynb}{Julia code for meshing of bi-material beam}. The problem consists of cooling the material from a reference temperature. The reference temperature has been taken as 300 $\tccentigrade$. The temperature of the domain is slowly reduced at each time step. Since we are reducing the temperature uniformly at each time step over the whole domain, we need not consider the temperature evolution equation. The resulting displacement can be directly obtained by using the governing equation corresponding to the displacement field. This is similar to the one adopted by Mandal {\it{et al.}} \cite{mandal2021fracture} wherein the load has been applied in the form of temperature strain based on the temperature change. The crack propagates because of thermal stresses due to the different coefficients of thermal expansion. Fig. \ref{fig:BiMatCrack} shows the crack path obtained in the domain. Also, Fig. \ref{fig:BimaterialComparison} shows the experimental crack pattern reported in \cite{grutzik2018crack} and the crack path obtained by this study near the crack tip. One can observe that the predicted crack path by the proposed model is  in good agreement with the experimental results. Prediction of crack path using the proposed model is similar to the prediction by Mandal {\it{et al.}} \cite{mandal2021fracture} that also does not show experimentally observed little drop in the crack path before going straight .

\subsection{Quenching test}
\begin{figure}[ht!]
 \centering
    \begin{subfigure}[b]{0.45\textwidth}
         \centering
 \includegraphics[width=1\textwidth,height=0.1\textheight]{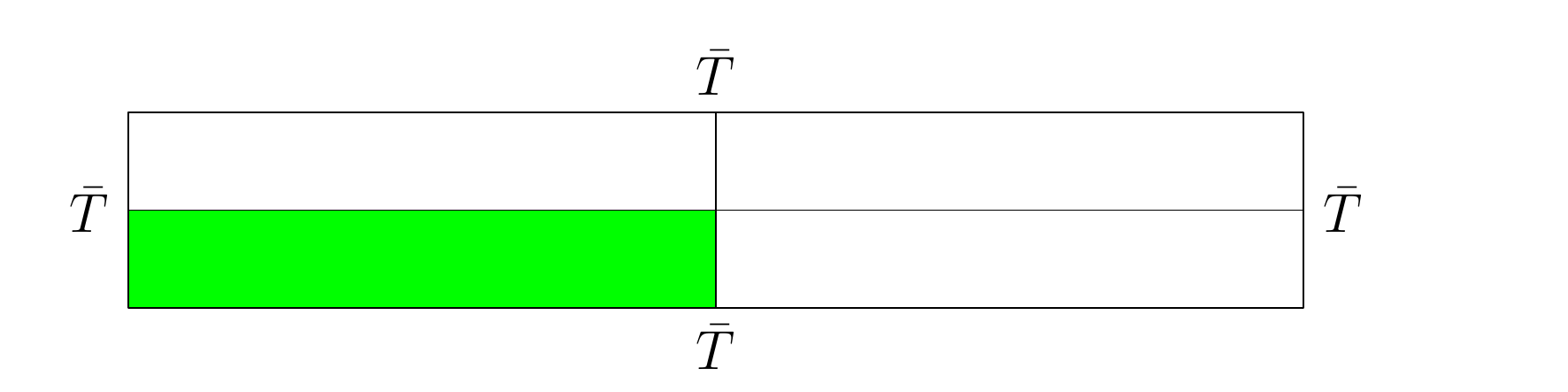}
         \caption{}
         \label{fig:QuenchingFullDomain}
     \end{subfigure}
    \begin{subfigure}[b]{0.45\textwidth}
         \centering
 \includegraphics[width=1\textwidth,height=0.085\textheight]{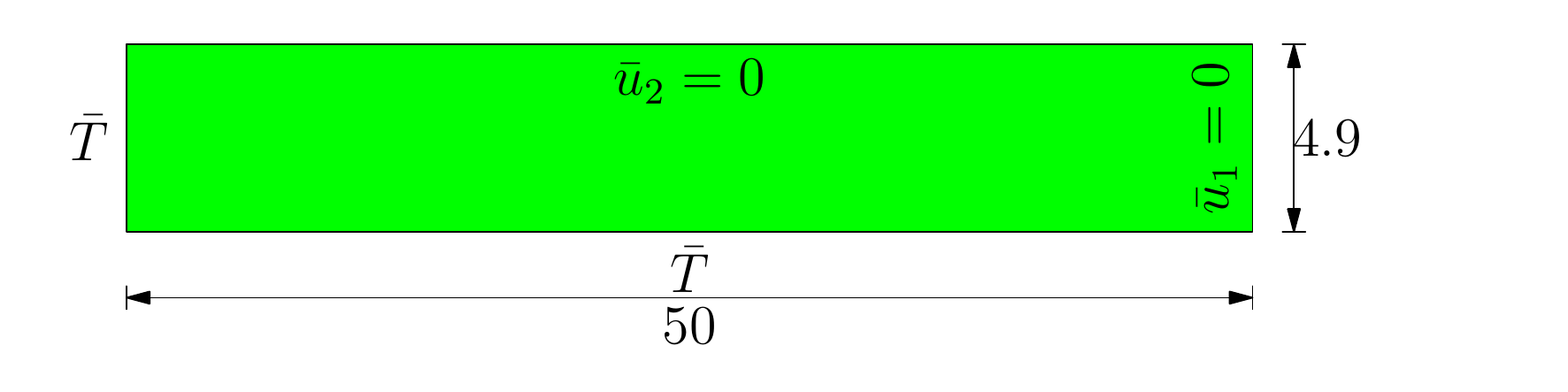}
         \caption{}
          \label{fig:QuenchingAnalyzedDomain}
     \end{subfigure}
  \caption{Geometry and boundary condition for quenching test: (a) complete domain (b) quarter of the domain with symmetric boundary conditions analyzed in the present simulation (dimensions are in mm).}
  \label{fig:QuenchingGeometry}
  \end{figure}

\begin{figure}[h]
     \begin{subfigure}{0.7\textwidth}
 \includegraphics[width=0.7\textwidth]{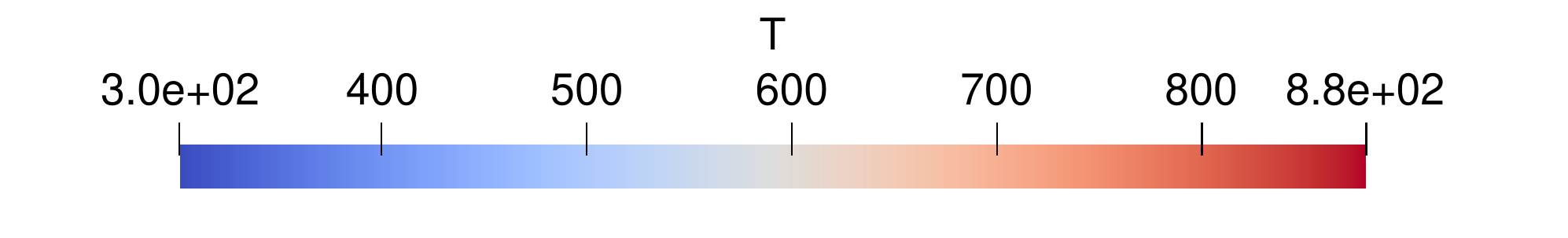}
     \end{subfigure}
     \begin{subfigure}{0.7\textwidth}
 \includegraphics[width=0.7\textwidth,height=0.06\textheight]{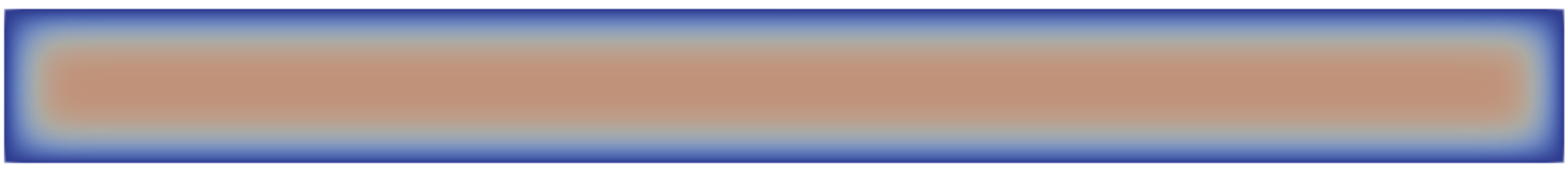}
 \caption{}
\label{fig:TempVariationTemp680}
     \end{subfigure}
      \begin{subfigure}{0.7\textwidth}
 \includegraphics[width=0.7\textwidth,height=0.06\textheight]{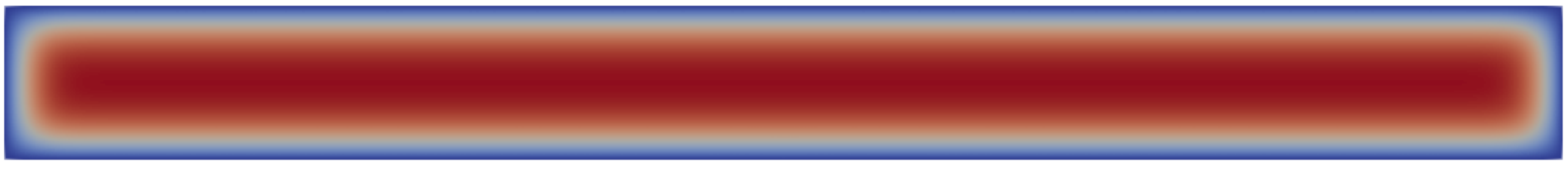}
\caption{}
\label{fig::TempVariationTemp980}
     \end{subfigure}
\caption{Variation of temperature field across the domain for (a) model with initial temperature 680 K and (b) model with initial temperature 880 K.}
\label{fig:QuenchingTempResults}
\end{figure}

\begin{figure}[ht!]
 \centering
     \begin{subfigure}[b]{0.45\textwidth}
         \centering
 \includegraphics[width=\textwidth,height=0.06\textheight]{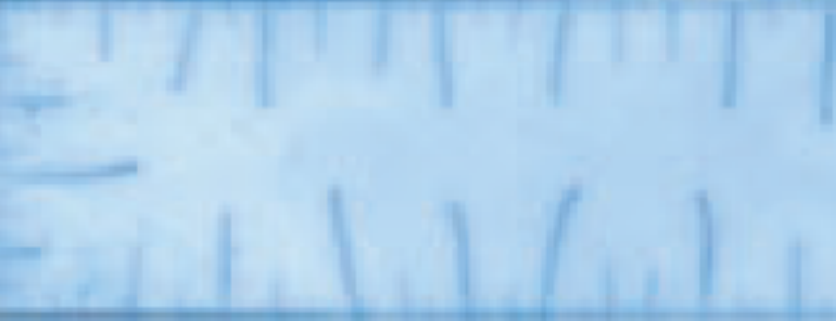}
         \caption{}
          \label{fig:QuenchExpTemp680}
     \end{subfigure}
      \begin{subfigure}[b]{0.45\textwidth}
         \centering
 \includegraphics[width=\textwidth,height=0.062\textheight]{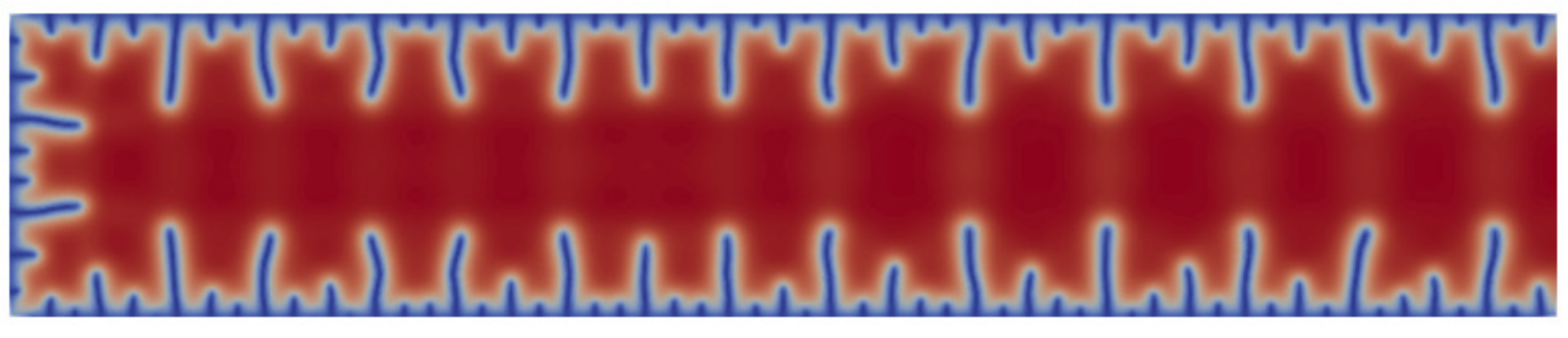}
         \caption{}
          \label{fig:PhaseFieldTemp680}
     \end{subfigure}
      \begin{subfigure}[b]{0.45\textwidth}
         \centering
 \includegraphics[width=\textwidth,height=0.06\textheight]{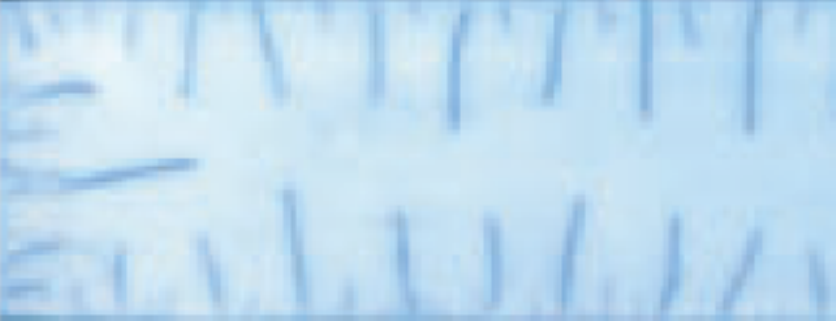}
         \caption{}
          \label{fig:QuenchExpTemp980}
     \end{subfigure}
      \begin{subfigure}[b]{0.45\textwidth}
         \centering
 \includegraphics[width=\textwidth,height=0.062\textheight]{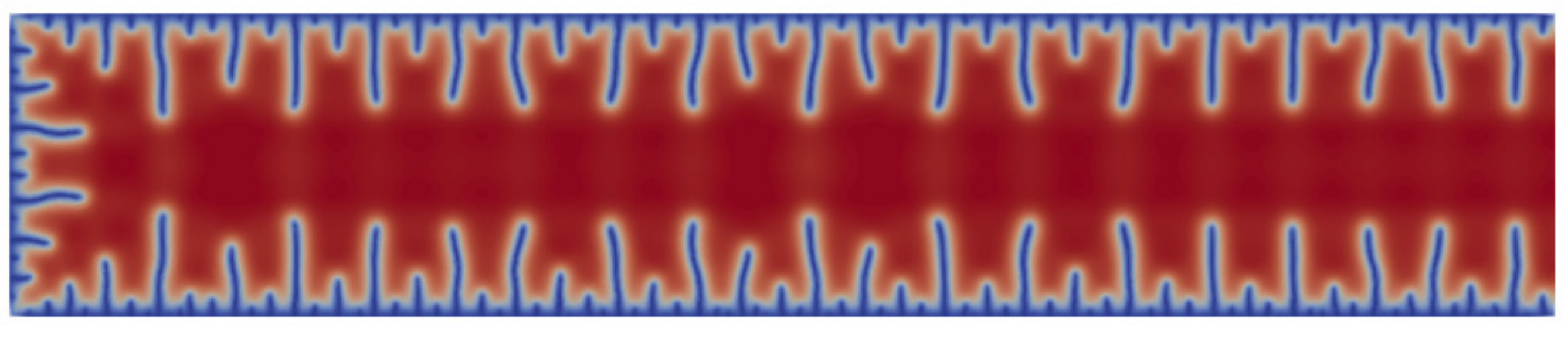}
         \caption{}
         \label{fig:PhaseFieldTemp980}
     \end{subfigure}
  \caption{Crack patterns in quenching tests: (a) experimentally observed crack pattern for a test with initial temperature 680 K, (b) predicted crack pattern by the proposed phase-field for a test with initial temperature 680 K, (c) experimentally observed crack pattern for a test with initial temperature 880 K, and (d) predicted crack pattern by the proposed phase-field model for a test with initial temperature 880 K.}
\label{fig:QuenchingResults}
\end{figure}

The experimental results generated by Shao {\it{et al.}} \cite{shao2011effect} for quenching test is considered as a benchmark problem for numerical modeling of thermo-mechanical problems. For the present case, we have considered a specimen with an initial temperature of $680$ K and $880$ K quenched into an ambient temperature of $300$ K. For the model development, we have considered only a quarter of the actual specimen due to the symmetry of the problem, as shown in Fig. \ref{fig:QuenchingGeometry} with a height of $4.9$ mm and length of $50$ mm. The horizontal and vertical displacements have been fixed on the right and top edges, respectively.  The bottom edge and right edge have been subjected to quenching through conduction by applying a fixed $300$ K temperature.  The length scale parameter has been selected as (length of the specimen)/250, and the element size has been selected such that $l_s$ is at least two times the mesh size. The Julia code used for creating the finite element mesh for quenching test specimen is given in \href{https://github.com/MdMasiurRahaman/PhaseFieldModelForThermoMechanicalFracture-/blob/main/MeshQuenchingTest.ipynb}{Julia code for meshing of quenching test specimen}. For the simulation, we have considered the same material properties as mentioned in Section \ref{sec:SENT} under plane stress condition similar to Mandal {\it{et al.}} \cite{mandal2021fracture}. We haven't considered the degradation of the $k$ considering the complex flow of water through the developed cracks for this problem. The distribution of temperature field across the domain is given in Fig. \ref{fig:QuenchingTempResults} and phase-field variable across the domain is demonstrated in Fig. \ref{fig:QuenchingResults} along with the experimentally found results by \cite{shao2011effect}. The results show that, as the initial temperature increases, i.e., for higher temperature difference, the crack density is higher, and the depth of crack propagation is also higher. The obtained results are found to be in good agreement with the experimental results of Shao {\it{et al.}} \cite{shao2011effect}, and numerical results by Wang {\it{et al.}} \cite{wang2020phase}, and Tangella {\it{et al.}} \cite{tangella2021hybrid}.
  
\section{Conclusions}
\label{sec:conclusion}
In this work, we have developed a thermodynamically consistent phase-field model for brittle fracture under thermo-mechanical loading. We have provided an open-source implementation of the developed model using the finite element toolbox, Gridap in Julia.  Our proposed model provides an equation for temperature evolution that can account for the effects of dissipative forces such as viscous damping whenever needed. We have validated the proposed model and its implementation against a few representative examples available in the literature. We have outlined all the steps to implement our proposed model in Julia for simulating fracture in a single edge notched plate that one can easily modify to generate results for other thermo-mechanical fracture problems.  Numerical results for a cruciform-shaped plate show the capability of our proposed model to reproduce the behavior under steady-state conditions. We have reproduced the effect of thermo-mechanical coupling on the global response (load-displacement curve) for a single edge notched plate under tension. The proposed model shows excellent agreement with the experimentally observed crack paths for a bi-material beam and the quenching test.  

Our proposed implementation is available with an open-source license. Hence, it eliminates the technical barrier for practitioners and researchers interested in exploring the phase-field model for solving a wide range of thermo-mechanical brittle fracture problems. Moreover, our proposed implementation will expose the users to many built-in packages of Julia that may be useful for researchers who want to extend the proposed implementation for the case of ductile fracture or other applications. The proposed multifield solution scheme can be implemented very easily for other coupled problems with the help of Gridap. Most importantly, the availability of an efficient and compact open-source code will provide better flexibility to the researchers and will establish a high standard for the research work.

\section*{Data accessibility} The present work does not generate any experimental data. Julia scripts used as the source codes to generate the numerical results presented in the manuscript can be downloaded as jupyter notebook files from \href{https://github.com/MdMasiurRahaman/PhaseFieldModelForThermoMechanicalFracture-}{Julia code for phase-field modeling of thermo-mechanical fracture}.

\section*{Declaration of competing interest}
Authors of this article declare that they have no conflict of interest.

\section*{Acknowledgments}
Authors of this article gratefully acknowledge support from the Indian Institute of Technology Bhubaneswar under the grant SP107.

\bibliographystyle{asmems4}

\bibliography{References}

\end{document}